\documentclass[preprintnumbers,amsmath,amssymb,aps]{revtex4}
\usepackage{graphicx}

\newcommand{\bee}{\begin{equation}}
\newcommand{\ee}{\end{equation}}
\newcommand{\beea}{\begin{eqnarray}}
\newcommand{\eea}{\end{eqnarray}}
\newcommand{\gfive}{\gamma_5}
\newcommand{\sign}{{\rm sign}}

\def\Tr{{\rm Tr}}

\begin{document}

\title{Chiral properties of two-flavor QCD in small volume and at large lattice spacing }

\author{Thomas DeGrand}
\author{Stefan Schaefer}
\affiliation{
Department of Physics, University of Colorado,
Boulder, CO 80309 USA
}

\begin{abstract}
We present results from simulations of two flavors of dynamical overlap fermions
on $8^4$ lattices at three values of the sea quark  mass and a lattice spacing of
about $0.16~{\rm fm}$. We measure the topological susceptibility and the 
chiral condensate.
A comparison of the low-lying spectrum of the overlap
operator with predictions from random matrix theory is made. To demonstrate the
effect of the dynamical fermions, we compare meson two-point functions with quenched
results.
Algorithmic improvements over a previous publication and
the performance of the algorithm are discussed.
\end{abstract}

\maketitle

\section{Introduction}

Chiral symmetry is a fundamental part of the theory of strong interactions and 
therefore should be respected when putting QCD on the lattice.
It was realized more than two decades ago, that this can be done 
with Dirac operators $D$ which (at zero quark mass) obey the Ginsparg--Wilson 
equation~\cite{Ginsparg:1981bj}
\bee
D\gfive +\gfive D = \frac{a}{R_0} D \gfive D 
\ee
where $a$ is the lattice spacing and $R_0$ the radius of the Ginsparg--Wilson circle.
A decade ago, this constraint was realized by overlap fermions 
\cite{Neuberger:1997fp,Neuberger:1998my}, fixed-point fermions
\cite{Hasenfratz:1993sp} and domain wall fermions \cite{Kaplan:1992bt,Furman:1994ky}.
Among those, only domain wall fermions have been used for some time in simulations
on the lattice which include the fermionic determinant. Only recently the first
steps toward dynamical simulations using overlap fermions have been taken 
\cite{Bode:1999dd,Fodor:2003bh,Fodor:2004wx,DeGrand:2004nq,Cundy:2005pi}. Because
of the high cost of applying the Dirac operator these are still limited to a small 
volume.
In this paper we present results from simulations with two flavors of dynamical overlap 
fermions in a small box and at a large lattice spacing.
We measure the topological susceptibility and, by comparing the low-lying eigenvalue
distribution to random matrix theory, the chiral condensate.
We also fix the lattice spacing and look for dynamical effects in meson two-point functions.
The measurement of the susceptibility is greatly facilitated as compared to, e.g.,
simulations using highly improved staggered fermions \cite{Aubin:2004qz}. 
With the overlap operator, we can use the same Dirac operator in the simulation 
and in the definition of the topological charge via the index theorem. 
The topological charge defined in this way thus directly influences the weight of a
configuration.

A more technical consequence of that is the  discontinuity of the 
fermionic determinant as a function of the gauge variables. The
surfaces in gauge field space at which the fermionic action is discontinuous
coincide with the change in topology defined by the index of the Dirac operator.
Although this does not pose a problem in principle, in practice changing the 
topological sector turns out to be a significant problem. As recent simulations
with highly improved staggered fermions show, this problem itself, however, is not
restricted to chiral formulations of QCD on the lattice \cite{Bernard:2003gq}.

In a recent paper \cite{DeGrand:2004nq}, we  described the setup of our simulations 
of two degenerate flavors of overlap fermions.
We studied the impact of fat (stout\cite{Morningstar:2003gk})
 links and multiple pseudo-fermion fields. 
Some improvements to the algorithm will be reported in the following.
The simulations presented in this paper used very
limited computer resources, i.e. half a year on an array of 12  3.2 Ghz Pentium-IVE's.
With that we present some results on $8^4$ lattices and a relatively coarse
lattice spacing of around $a=0.16~{\rm fm}$ and a quark mass  down to
about $35~{\rm MeV}$.

This paper is organized as follows: We first review the algorithm,  describe
the improvements over our previous publication and  give the parameters of our simulation.  
Then in Sec.~\ref{sec:R0} we attempt to set the lattice spacing by determining the Sommer parameter $r_0$ and 
proceed with the extraction of the topological susceptibility in Section~\ref{sec:susc}. In 
Section~\ref{sec:twopt} 
we look at meson two-point functions. By comparing
to quenched results on matched lattices, we demonstrate the impact of dynamical fermions on the 
scalar correlator.
Finally, we compare the low-lying spectrum of the overlap operator with the predictions from
random matrix theory.
A note in caution: for all these investigations, the volume which we are simulating is  too small.
Future simulations will improve on this. Here we want to demonstrate that simulations with
dynamical overlap fermions are possible and give results which match  
our expectations of full QCD in a small box.

\section{Definitions and Algorithm\label{sec:alg}}

Let us fix the conventions and describe the algorithm together with the improvements 
since our previous publication~\cite{DeGrand:2004nq}, to which we refer the reader
for more details.
The massive  overlap operator is given by \cite{Neuberger:1997fp,Neuberger:1998my} 
\bee 
D_{ov}(m)= (R_0-\frac{m}{2}) \left[ 1+\gfive \epsilon(h(-R_0))\right]+m
\label{eq:Dov}
\ee 
with $\epsilon(h)=h/\sqrt{h^2}$ the sign function of the Hermitian kernel 
operator $h=\gfive d$ which is taken at the negative mass $R_0$.  
$R_0$ is  the radius of the Ginsparg--Wilson circle. We are 
using a planar kernel Dirac operator $d$ with nearest and next-to-nearest (``$\sqrt{2}$") interactions. 
For details see Ref.~\cite{DeGrand:2004nq}. The sign function is computed using the
Zolotarev approximation 
with an exact treatment of the low-lying eigenmodes $|\lambda\rangle$ of $h(-R_0)$
\bee
\epsilon(h(-R_0)) = \sum_i \sign (\lambda_i) | \lambda_i \rangle \langle \lambda_i |
\approx \sum_{i=1}^{n_{\rm eig}} \sign (\lambda_i) | \lambda \rangle \langle \lambda |+
h \sum_j \frac{b_j}{h^2+c_j}(1-\sum_{i=1}^{n_{\rm eig}} |\lambda_i\rangle\langle \lambda_i|)\ . \label{eq:sign}
\ee

For our simulation we use the Hybrid Monte Carlo (HMC)  algorithm \cite{Duane:1987de} as
modified for overlap fermions by Ref. \cite{Fodor:2003bh}.
The effective action is given by 
\bee
S_{\rm eff} =\phi_0^+ H^{-2}(m_0) \phi_0
+ 
\sum_{i=1}^{n_{pf}-1} \phi_i^+ H^{-2}(m_i) H^2(m_{i-1}) \phi_i 
+ 
S_g[U]
\label{eq:effact}
\ee
with $H^2(m)=D_{ov}(m)^+D_{ov}(m)$ the square of the hermitian overlap Dirac 
operator. $S_g[U]$ is the gluonic action. The $\phi_i$ are the $n_{pf}$ pseudo-fermion
fields used to include the contribution of the fermion determinant. The use of 
several of these fields has been suggested in \cite{Hasenbusch:2001ne,Hasenbusch:2002ai}.
It improves the stochastic estimate of the determinant. We studied its effects extensively
in Ref.~\cite{DeGrand:2004nq}.

With the HMC algorithm, an ensemble distributed according to this effective action is
generated by updating the gauge fields using molecular dynamics trajectories with
a final accept-reject step. At the beginning of the trajectory one chooses
momenta $\pi$ conjugate to the gauge fields and also refreshes the pseudo-fermions.
One then integrates the resulting equations of motion (treating $S_{\rm eff}$ as the potential)
numerically in some fictitious simulation time $\tau$. They result from the requirement
that the total `energy' ${\cal H} = \pi^2/2+S_{\rm eff}$ is conserved. At the end one applies an 
accept/reject step with acceptance probability $P=\min[1,\exp(-\Delta {\cal H})]$ which 
corrects for the errors in the numerical integration.

An important difference between conventional fermions and overlap fermions is 
that the effective action $S_{\rm eff}$ is discontinuous for the latter. 
The discontinuity has its origin in the sign function in the definition of the Dirac
operator Eqs.~(\ref{eq:Dov}) and (\ref{eq:sign}).
The fermionic action is discontinuous at the surfaces
where the kernel operator $h(-R_0)$ has a zero eigenvalue. According to the index
theorem~\cite{Hasenfratz:1998ri}, these are also the surfaces 
where the topological charge (as seen by the fermions) changes. Ref.~\cite{Fodor:2003bh}
gives the prescription for how to deal with this situation. The molecular dynamics evolution 
can be thought of as resembling that of a classical particle in the presence of a potential step. 
If the step in the action is too big for the particle to get across it, the particle is
reflected, i.e. the momentum component normal to the zero eigenvalue surface is reversed.
On the other hand, if the normal component is large enough to change topological sector,
the normal component is reduced such that energy is conserved. Following 
Ref.~\cite{Fodor:2003bh} this is called a refraction. The momentum is then altered 
according to
\bee
\Delta \pi =
\begin{cases}
-N \; \langle N |\pi \rangle + N \; \sign  \langle N | \pi \rangle \; 
 \sqrt {\langle N | \pi \rangle^2-2 \Delta S_f}
& \text{if  $\langle N | \pi \rangle^2>2 \Delta S_f $}\\
-2  N \langle N | \pi \rangle &  \text{if $\langle N | \pi \rangle^2\leq 2 \Delta S_f$}
\end{cases}
\label{eq:ref}
\ee
with $N$ the vector normal to the zero eigenvalue surface, $\pi$ the momentum and $\Delta S_f$
the discontinuity in the fermionic action.
To monitor whether an eigenvalue has changed sign, one thus has to compute some number 
$n_{\rm eig}$ of the lowest eigenmodes of $h(-R_0)$
and see whether the eigenvalue of any of them changes sign. The matching is done
by computing the scalar products  of the low modes before and after a step.  Since the eigenmodes
are needed  anyway to precondition the construction of the sign function, there is virtually 
no overhead associated with this test.

Note that this part of the algorithm potentially scales with the square of the 
volume: The cost of determining the height of the step is at least proportional to the volume. 
The number of times this procedure has to be executed can be assumed to be proportional 
to the density of eigenmodes of the kernel operator at the origin, which in turn might be 
proportional to the volume. It is therefore pivotal to keep the cost of this 
step as low as possible.

A major improvement in the algorithm 
is the way in which we compute the height of the step. In our previous publication,
 we ran two conjugate gradients to compute
\bee
\Delta S = \Delta \left [ \phi_0^+ H^{-2}(m_0) \phi_0
+
\sum_{i=1}^{N-1} \phi_i^+ H^{-2}(m_i) H^2(m_{i-1}) \phi_i
\right]
\ee
where the difference is taken between the fermionic actions for which only the 
sign of the lowest eigenmode (the one which becomes zero on the surface) is changed without
changing any of the other modes or the gauge configuration. This was very expensive since we
have to decide frequently whether to refract or reflect.
Because the square of the hermitian Dirac operator in one chiral sector $\sigma=\pm 1$ is  of the form
\bee
H^2_\sigma(m)= 2 (R_0^2-\frac{m^2}{4})
P_\sigma\left[1+\sigma \sum_i \epsilon(\lambda_i) |\lambda_i\rangle\langle \lambda_i| \right]P_\sigma+m^2
\label{eq:shift}
\ee
with $|\lambda\rangle$ the eigenvectors of $h(-R_0)$ and $P_\sigma=\frac{1}{2}(1+\sigma \gfive)$ the projector
on the chiral sector,
the sign change of the crossing mode amounts to the change 
\bee
H^2_\sigma(m) \longrightarrow H^2_\sigma(m)\pm(4R_0^2-m^2) P_\sigma|\lambda_0\rangle\langle \lambda_0|P_\sigma
\ee
with $|\lambda_0\rangle$ the zero mode and  the sign being minus the product of the 
sign associated with the chiral sector and the sign of the eigenmode before the step.
Thus, we can use the Sherman-Morrison formula~\cite{Golub} to compute the height of the step.
The result is
\bee
\Delta \left [ \langle \phi| P_\sigma \frac{1}{H_\sigma(m)^2}P_\sigma| \phi \rangle \right] = 
\mp 
\frac{(4R_0^2-m^2)}
{1\pm (4R_0^2-m^2) \langle \lambda_0|P_\sigma H^{-2}_\sigma(m) P_\sigma|\lambda_0 \rangle}
|\langle  \phi |P_\sigma\frac{1}{H_\sigma(m)^2}P_\sigma| \lambda_0\rangle|^2 \ .
\label{eq:sherman}
\ee
This has the additional advantage that instead of running two conjugate gradients for each 
pseudo-fermion field, we only have to invert once, using the eigenmode which changes sign
as a source. Because
the height of the step is directly computed, one also has better control over the accuracy --- 
compared to taking the difference of two approximate quantities. 

One can further exploit Eq.~\ref{eq:sherman}
 by realizing that the  cost the of the inversion of $H^2(m)$ depends on the 
chirality of the source and the topological sector in which one is inverting.
One can compute the
height of the step from either side of the surface. In the chiral sector with
 zero-modes, it is cheaper to invert in the topological sector of lower charge. However, since
the zero-modes push the spectrum of the other modes up, 
the conditioning number of $H^2_\sigma(m)$ in the sector 
without zero-modes is lower on the side of the surface with higher topology.
A second advantage of the use of this formula is that we can
 monitor the step-height during the CG iterations and terminate the iteration when it has
become clear that we are going to reflect.

Finally, let us report a small improvement in the computation of the fermion force $\delta S_f[U]$.
The derivative of the Zolotarev part of the  approximation to the sign function Eq.~(\ref{eq:sign})
has been given in many places. For each pseudo-fermion field and each order in the rational approximation
the formula has a term $1/(h^2-c)$, the derivative of which is
 \bee
-\frac{1}{h^2-c} (h\delta h+\delta h h) \frac{1}{h^2-c}.
\ee
One thus has to invert the kernel action, which can be done simultaneously for all shifts
using a multi-mass algorithm. The computation of $\delta h$ follows via standard 
methods from Ref.~\cite{Gottlieb:1987mq}. However, due to the many shifts in the Zolotarev approximation
and several pseudo-fermion fields, this part is a non-negligible contribution to the total cost
of the simulation.

More difficult is the derivative of the projector term $P_\lambda=|\lambda\rangle\langle \lambda|$. 
This derivative is basically given by first-order perturbation theory (See Ref.~\cite{Narayanan:2000qx})
\bee
\delta P_\lambda=\frac{1}{\lambda-h}(1-P_\lambda)\delta h P_\lambda+
P_\lambda \delta h^+\frac{1}{\lambda-h}(1-P_\lambda) .
\ee
 Because $(\lambda-h)$ is singular, its inversion 
is problematic even though the contribution of the eigenmode with eigenvalue $\lambda$ is projected out
of the source:
since both $\lambda$ and $|\lambda\rangle$ are only approximately known, one faces a 
``zero divided by zero'' problem. In our previous
publication, we therefore shifted the pole position, performed the inversion,
 and interpolated our result. This turned out to be insufficiently stable.
Now we use a Chebychev approximation to the inverse of $h^2-\lambda^2$ in the range
such that the inverse is precise outside the known eigenvalues of $h(-R_0)$. The 
advantage is that this approximation is finite at $h^2-\lambda^2=0$. The problem is thus
reduced to a contribution from the eigenvalue $\lambda$ mode which is
  zero times some finite number given by the required accuracy and 
the range in which one computes the eigenvalues explicitly.

Eq. \ref{eq:shift} provides us with a tantalizing result,
 which we do not know how to apply in the context of HMC: an exact formula for the ratio
 of the fermion determinants on either side of the topology-changing boundary:
\bee
  \frac{\det \tilde H^2_\sigma(m)}{\det  H^2_\sigma(m)}
 =  1\pm(4R_0^2-m^2) \langle \lambda_0 |P_\sigma \frac{1}{H^2_\sigma(m)} P_\sigma | \lambda_0\rangle \ .
 \label{eq:stepdet}
\ee
HMC does not ever use the exact determinant as part of the simulation. Instead,
it generates configurations whose statistical weight is controlled by
the effective action Eq. \ref{eq:effact}. In an algorithm which does approximate the determinant
directly, like the $R_0$ algorithm, it seems straightforward; one would just use the 
logarithm of Eq. \ref{eq:stepdet} as the step.
However, we are unwilling to abandon HMC for two reasons: First,
we feel that we benefit substantially
from the fact that HMC is exact and therefore do not want to run an $R$ type algorithm which cannot 
be made exact in an obvious way.  Second, we gain considerable speed using HMC over an R algorithm
because in HMC we can use a previous solution to an inverse of $H(m)^2$ to
 begin the computation of the new force.
This point deserves further investigation.

\section{Simulation Parameters and Performance of the Algorithm}

We simulate on $8^4$ lattices at one value of the gauge coupling $\beta=7.2$
which we chose to be roughly at the $N_t=6$ phase transition (for and 
overview over the lattice spacings see Table~\ref{tab:R0}). We use the L\"uscher--Weisz gauge 
action~\cite{Luscher:1984xn} with the  tadpole improved coefficients of 
Ref.~\cite{Alford:1995hw}. Instead of determining the fourth root
of the plaquette expectation value $u_0=(\langle U_{pl}\rangle/3)^{1/4}$ 
self-consistently, we set it to 0.86 for all our runs as we did in our previous 
publication.

Our kernel operator $d$ is constructed from gauge links to 
which two levels of isotropic stout blocking~\cite{Morningstar:2003gk} have
been applied. The  blocking parameter $\rho$ is set to 0.15. 

We report on simulations at three values of the bare sea quark mass $am_q= 0.03$, $0.05$ and $0.1$.
Based on measured lattice spacings from the Sommer parameter and the perturbative calculation
of matching factors reported in the Appendix,
we believe that these values correspond to $\overline{MS}$ quark masses of about 35, 55 and 100 MeV.
An overview of our collected statistics is given in Table~\ref{tab:stat}. The trajectories
all have length one, divided in 20 elementary time steps.
 We use a Sexton-Weingarten \cite{Sexton:1992nu} integration scheme in order apply a smaller 
time step for the gauge field integration. We perform 12 applications of the  gauge force and
gauge field update per elementary time step. Again, see Ref.~\cite{DeGrand:2004nq} for details.

In order to monitor whether an eigenvalue has changed sign, we compute the lowest $8$ eigenmodes
of $h(-R_0)$ in each step. These are also used to precondition the construction of the sign 
function.

\begin{table}
\begin{tabular}{c|c|c|c|c|c|c}
$m_q$ \  &\  stream\  &\ $n_{\rm pf}$\  &\ trajects.\ &\ acc. rate\ &\ reflects\ &\ refracts\ \\
\hline
0.03  &  1     & 3   & 354  & 73\%  & 1658  & 31 \\ 
0.03  &  2     & 3   & 249  & 74\%  & 1136  & 35 \\ 
0.03  &  3     & 2   & 220  & 77\%  & 946   & 19 \\ 
0.03  &  4     & 3   & 173  & 68\%  & 891   & 19 \\ 
0.05  &  1     & 3   & 246  & 63\%  & 1350  & 98 \\ 
0.05  &  2     & 3   & 225  & 63\%  & 1033  & 51 \\ 
0.05  &  3     & 2   & 251  & 71\%  & 1089  & 36 \\ 
0.05  &  4     & 2   & 250  & 64\%  & 1233  & 68 \\ 
0.05  &  5     & 2   & 250  & 69\%  & 1218  & 43 \\ 
0.10  &  1     & 3   & 196  & 59\%  & 1043  & 113 \\ 
0.10  &  2     & 3   & 153  & 49\%  & 1040  & 153  \\ 
0.10  &  3     & 2   & 285  & 74\%  & 1334 &120  \\ 
0.10  &  4     & 2   & 191  & 71\%  & 945  & 95  \\ 
\end{tabular}
\caption{\label{tab:stat}Overview over our collected data at $\beta=7.2$ 
with several streams per quark mass. We give the number
of pseudo-fermion fields $n_{\rm pf}$ used, the number of trajectories, the 
acceptance rate and the total number of reflections and refractions. In our 
analysis we typically discard the first 100 configurations per stream. The
statistics includes all (accepted and rejected) trajectories. Note that
the number of refractions is higher than the number of effective changes
in topologies because of tunneling back and rejections.}
\end{table}

\begin{figure}
\includegraphics[width=0.3\textwidth, angle=-90, clip]{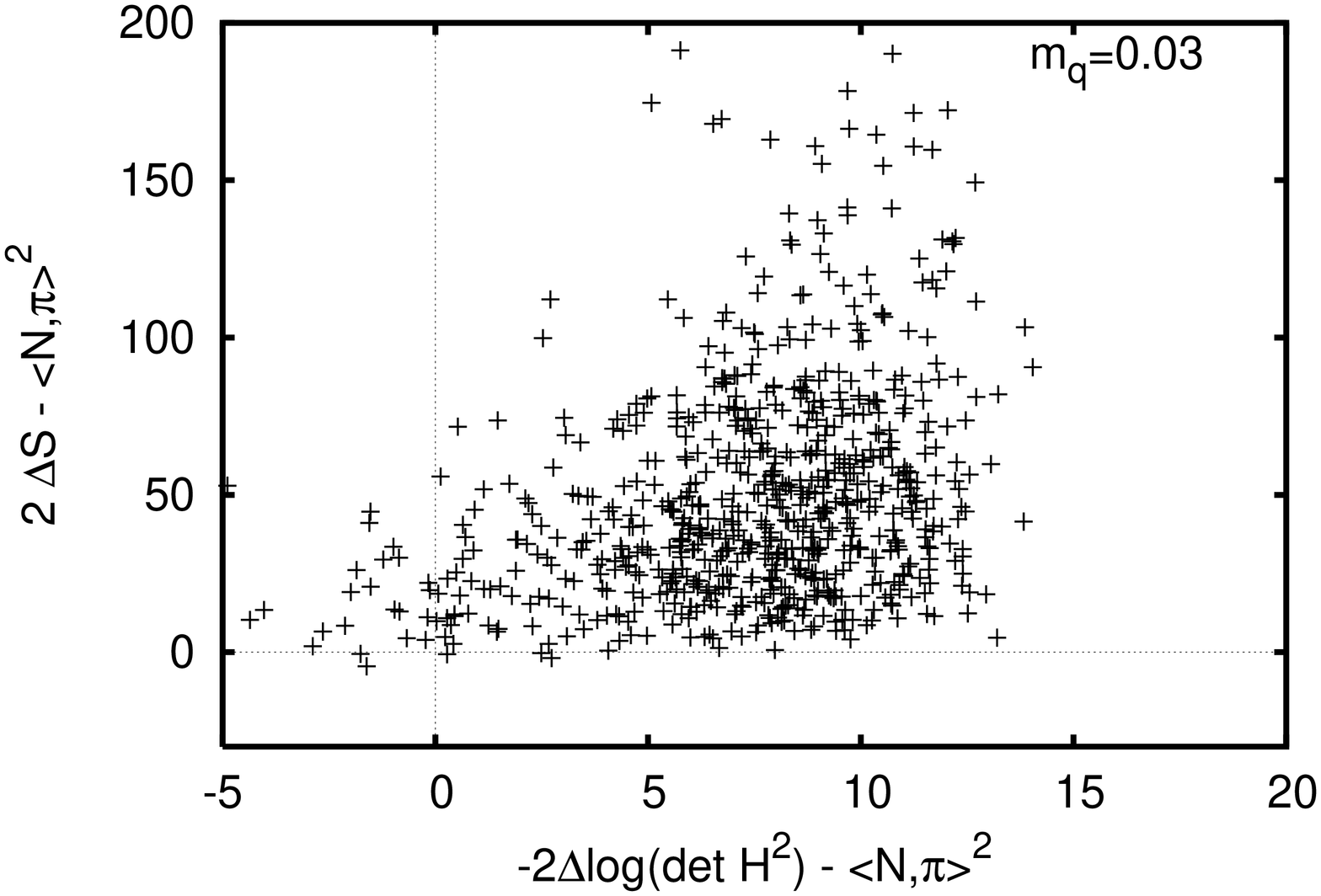}
\includegraphics[width=0.3\textwidth, angle=-90, clip]{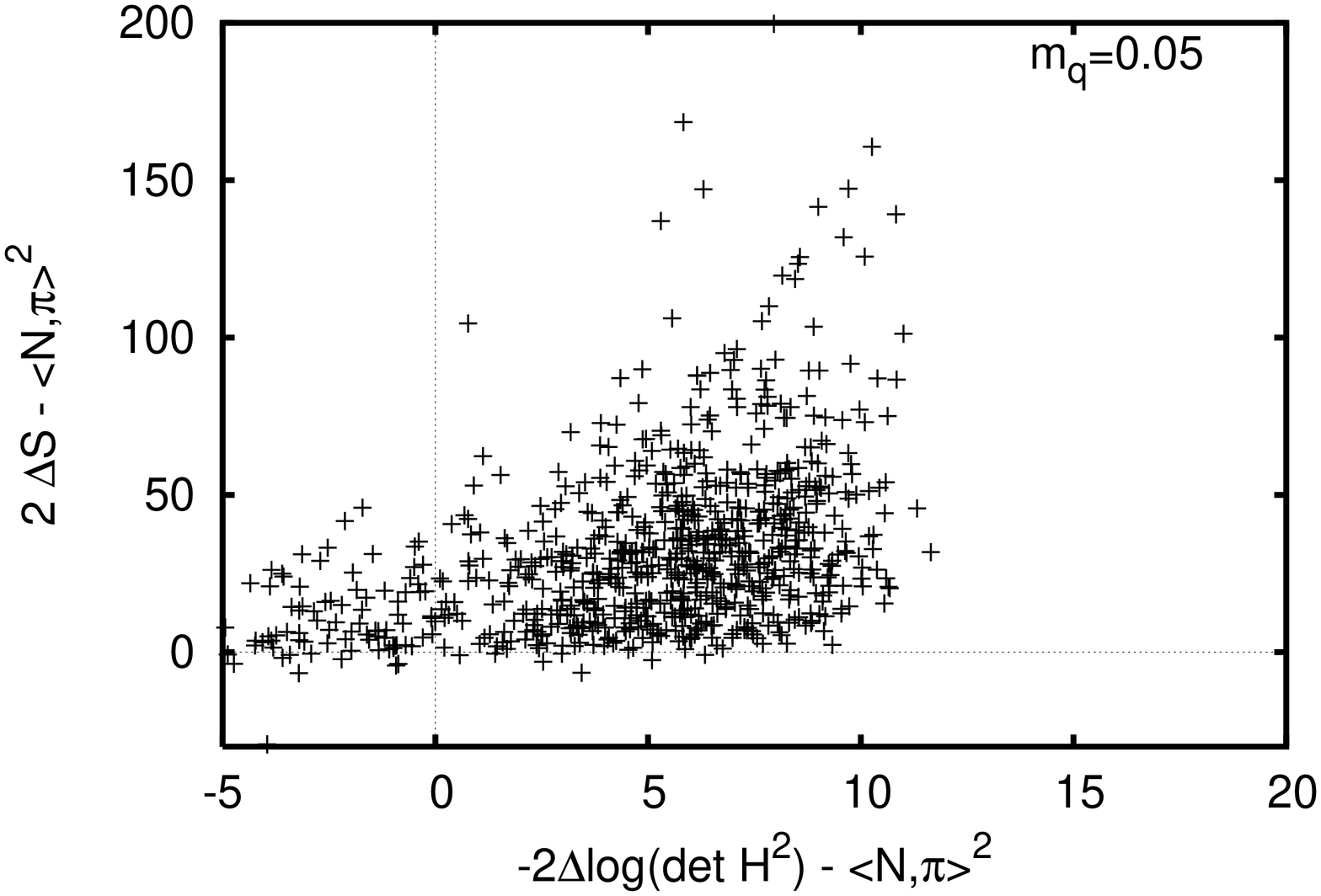}
\caption{The stochastic estimate of the height of the step compared to the actual change
in the logarithm of the determinant from a subset of our ensemble.
We subtracted the normal component of the momentum squared (which is typically less than 10) such 
that negative values mean refraction and positive ones reflection. For mass $m_q=0.03$ on
the left we have a number of events in the upper left quadrant that would have tunneled with the
exact change of the determinant and only a few that actually tunneled (in the two lower
quadrants). For $m_q=0.05$ the picture is similar, even though there are more tunneling events. 
\label{fig:1}}
\end{figure}

For our analysis, we typically discard the first 100 configurations in a stream and separate two
consecutive measurements by 5 trajectories. The separation of the configurations is based on
our measurement of the auto-correlation time of the plaquette. It is around $5(1)$ varying little
with the quark mass.
We observed no significant difference in acceptance rate between the quark masses.
However, the auto-correlation time of the topological charge differs enormously. This is due
to the fact that the estimation of the step height of the fermionic action intrinsic to HMC
is much more subject
to fluctuation
for smaller masses than for larger ones.
 As argued in our previous publication, a poor estimator
results in reflections. We partially address this problem by the use of multiple pseudo-fermion
fields, but the problem remains.
 In Fig.~\ref{fig:1} we show a scatter plot of the real change in the determinant at the
step, from Eq. \ref{eq:stepdet},
as compared to the stochastic estimate with three pseudo-fermion fields.
We subtracted the normal
component $\langle N, \pi \rangle^2$ of the momentum so that, according to Eq.~(\ref{eq:ref}),
negative values allow refractions. However, $\langle N, \pi \rangle^2$ is almost always smaller than 10. 
The stochastic estimate of the step height has a wide spread but is typically large.
Since $\exp(-\Delta S)$ will average to the ratio of the fermionic determinants on both sides of the step, 
this is a consequence of the large fluctuation in the estimator. (A few small values of $\Delta S$ 
have to be compensated by a large number of large ones, for which $\exp(-\Delta S)$ is approximately zero.)

\begin{figure}
\includegraphics[width=0.33\textwidth, angle=-90,clip]{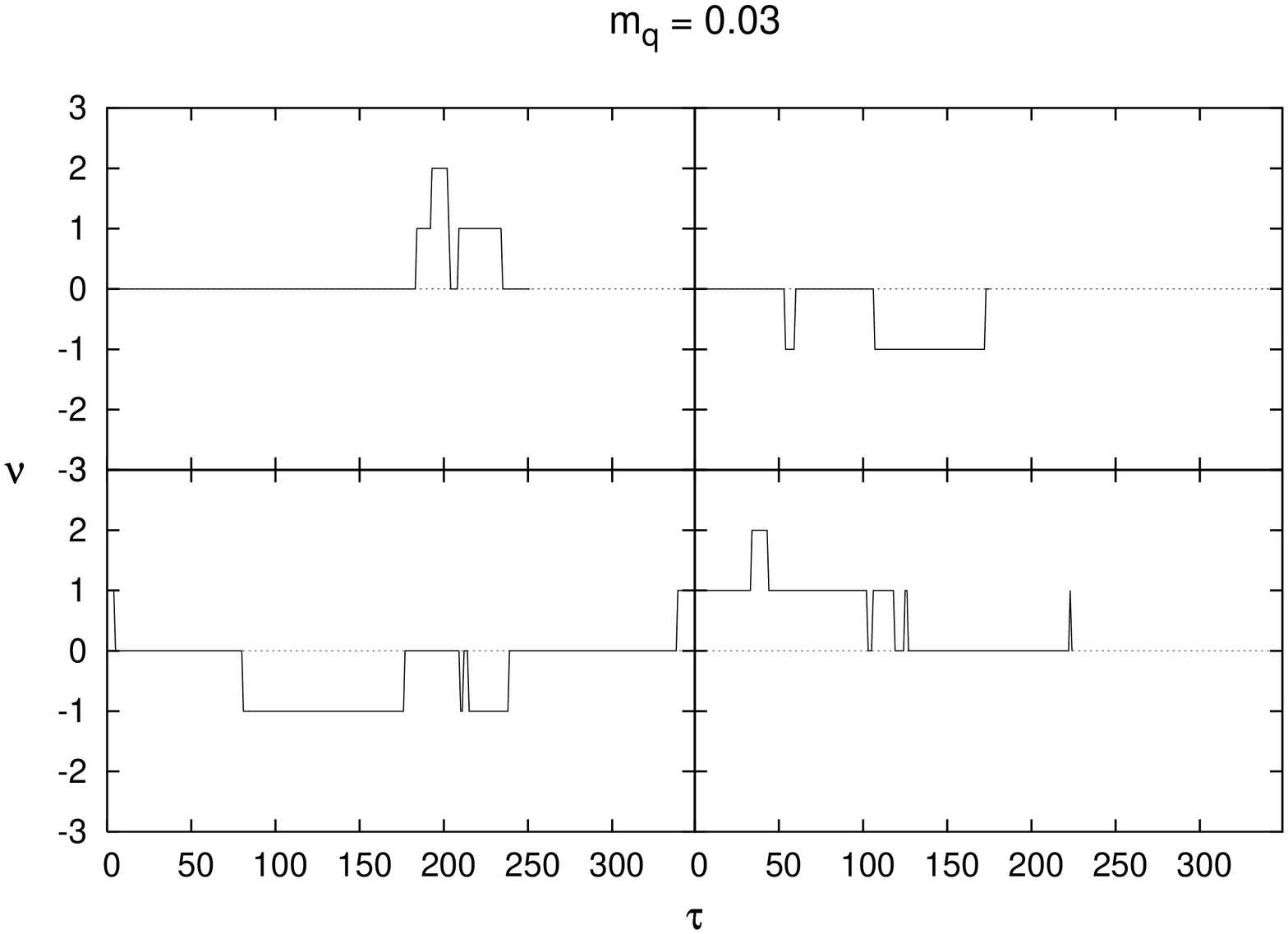}
\includegraphics[width=0.33\textwidth, angle=-90,clip]{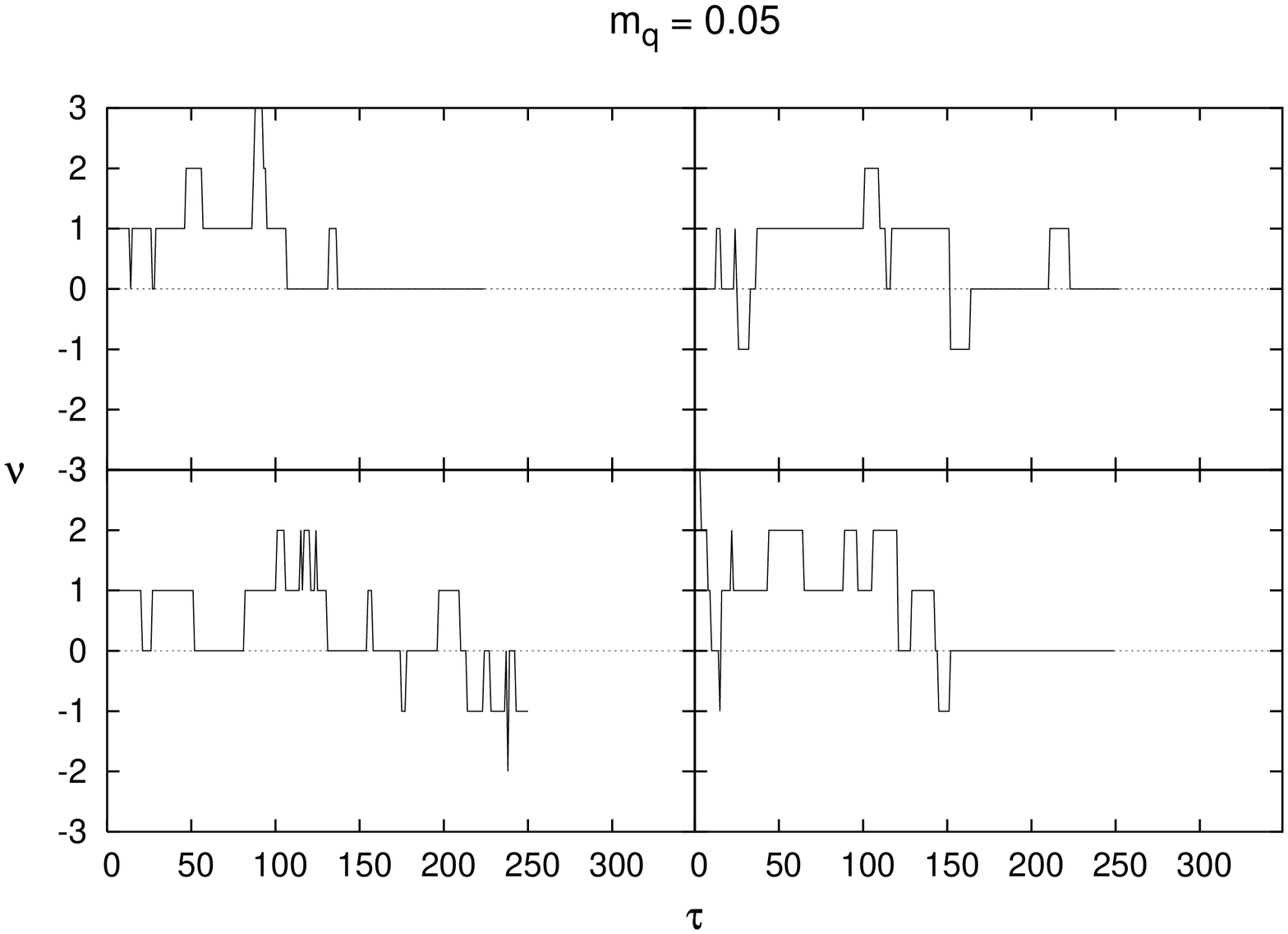}
\includegraphics[width=0.33\textwidth, angle=-90,clip]{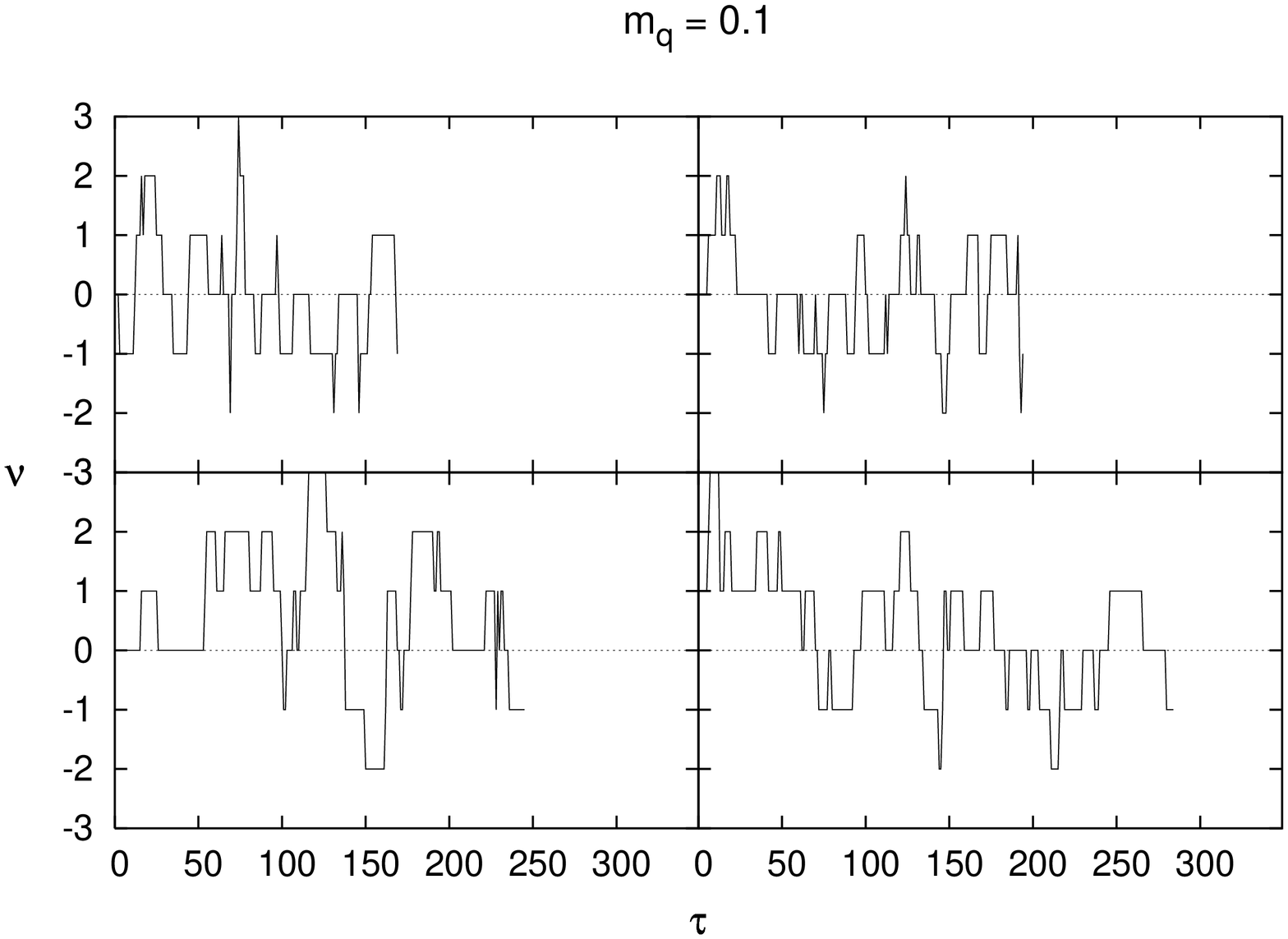}
\caption{\label{fig:topohist}The history of the topological charge $\nu$ as given by the index theorem for
our three quark masses. The data shown includes the thermalization runs.}
\end{figure}

The low correlation between the estimator and the physical step height Eq.~(\ref{eq:stepdet})
shows  up in the large auto-correlation time of the topological charge, whose time history is shown in
Fig.~\ref{fig:topohist}. Even
though part of it is physics --- lighter quarks make it harder to get from,
e.g.,  $\nu=0$ to $\nu=\pm 1$ --- the height of the step grows with $1/m^2$ instead 
of the expected determinant ratio, $\log~m$. Since the normal component of the momentum
is roughly independent of the quark mass, it becomes more and more difficult to change topology 
(also see discussion in Sec.~\ref{sec:susc}).
The large auto-correlation time for the topology is
a phenomenon  that is also known with other fermions, e.g. improved  staggered quarks. 
To the extent that these formulations know about topology, the step in the fermion action
for the overlap might be replaced for them by a steep region which approximates the step. 
The result is the same: if the approximation of the determinant is bad, the step is 
overestimated most of the time and one does not change topology.

Let us finally take a look at the relative cost of the various ingredients of the algorithm.
In Table~\ref{tab:time} we list the cost per call and the fraction of the total cost of
each of the major parts of the program. 
The conjugate gradient inversion of $H_{\rm ov}^2$  needed for the computation of the force,
starting action and the reflections/refractions takes by far the largest fraction.
Even though one inverts only on one source (the zero-mode of $h(-R_0)$)
in the reflection/refraction routine, these inversions are very expensive since there is no good 
starting vector. Therefore, they alone take a fifth to a quarter of the total cost,
depending on the quark mass. The inversions can be cheaper for the lighter quark mass because
there is less topology; the conditioning number of $H_{\rm ov}$ for $\nu=0$ is lower than for $|\nu|>0$.

The cost of computing the $8$ eigenvectors of the kernel operator is small, about $10\%$.
It is cheap because the eigenmodes of the kernel do not change much during the 
evolution; one thus has good starting vectors. We
need them to precondition the construction of the overlap operator and to monitor whether
an eigenvalue has changed sign. Finally, the computation of the
fermion force (outside the inversion of $H_{\rm ov}^2$) takes about a sixth of the 
total time. This is due to the inversion of the kernel operator, the computation of 
$\psi_i^+ \delta h(-R_0) \psi_i$ for each of the poles in the Zolotarev approximation
of the sign function and the inversion of $h(-R_0)$ for the projector term in the 
sign function as discussed at the end of Section~\ref{sec:alg}.

\begin{table}
\begin{tabular}{r|c|c|c|c}
&\multicolumn{2}{c|}{ $m_q=0.03$} &\multicolumn{2}{c}{ $m_q=0.05$}\\
&\ sec. per call\ &\ \% of total time\
&\ sec. per call\ &\ \% of total time\ \\
\hline
CG in MD evolution    \   &   406 & 48\% &  534 & 46\% \\
reflect. / refract.   \   &   800 & 18\% & 1029 & 26\% \\
fermion force         \   &   148 & 17\% &  175 & 15\% \\
eigenvalues of $h(-R_0)$\ &   110 & 12\% &  115 &  9\% \\
       setup source    \  &   577 & 3\%  &  634 &  3\% \\
       gauge update    \  &   27  & 3\%  &  27  &  2\% \\
\end{tabular} 
\caption{\label{tab:time}The cost in wall clock time for various parts of the algorithm
measured on a subset of the data.
The inversion of $H^2_{\rm ov}$ takes the largest part. It is responsible for the majority
of the cost in the first, second and fifth row, together about $75\%$.}
\end{table}

\section{Setting the scale\label{sec:R0}}

\begin{figure}
\includegraphics[width=0.33\textwidth, angle=-90,clip]{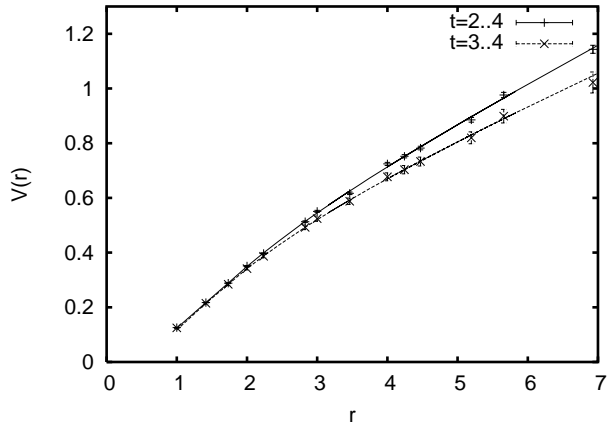}
\caption{\label{fig:3}The heavy quark potential extracted from fits to Wilson loops constructed
from HYP links.
The two sets result from two different fit ranges; $t\in [2,4]$ and $t\in [3,4]$. The curves are
fits of the form Eq.~(\ref{eq:pot}) to the potential which include a perturbative correction 
to account for the HYP links.}
\end{figure}

In order to get an idea at which lattice spacing we are simulating, we  measured 
 the heavy quark potential and
extracted the Sommer parameter $r_0$~\cite{Sommer:1993ce}. (It is defined as the 
distance at which the force between two heavy quarks $F(r)$ with distance $r$
satisfies $r_0^2F(r_0)=1.65$.) This is necessary because
we expect a substantial shift due to the dynamical fermions \cite{Hasenfratz:1993az}
with respect to the quenched lattice spacing~\cite{Gattringer:2001jf}.
Unfortunately, our lattice is relatively small. It is thus not possible to estimate
the uncertainty in the extraction of the potential from the data alone. 
To illustrate this we show in 
Fig.~\ref{fig:3} the potential extracted from  Wilson loops $W[r,t]$ which have 
been constructed from HYP smeared links\cite{ref:HYP}. The two sets correspond to single
exponential fits to the $W[r,t]$  with $t\in[2,4]$ and $t\in[3,4]$, respectively. 
The latter set gives a significantly smaller potential. 
The curves represent fits of the form
\bee
V(r) = \frac{A}{r}+B r+C+D f(r)
\label{eq:pot}
\ee
where $f(r)$ is a perturbatively determined correction~\cite{Hasenfratz:2001tw} to account 
for the effect of the HYP links on short distances. The results are given in Tab.~\ref{tab:R0}.

\begin{table}
\begin{tabular}{c|c|c|c|c}
$\beta=7.2$& \multicolumn{2}{c|}{ $t\in[3,4]$} & \multicolumn{2}{c}{  $t\in[2,4]$} \\
\hline
$m_q$      &     $r_0/a$     &      $a$[fm]     &     $r_0/a$     &     $a$[fm]     \\
\hline
0.03  \ \  & \ 3.27(3)  \  & \ 0.153(2) \    & \ 3.04(2) \     & \ 0.165(1) \   \\
0.05  \ \  & \ 3.17(5) \   &   0.158(2)       &  2.97(3)       &  0.168(1)       \\
0.10  \ \  & \ 3.00(4) \   &   0.167(2)       &  2.82(3)       &  0.177(2)         
\end{tabular}
\caption{\label{tab:R0}The Sommer parameter $r_0$ and the lattice spacing $a$ in fm from $r_0=0.5~{\rm fm}$.
We show results from two fit ranges used to extract the potential. The fit range for the fit to the
potential was $r\in[1.4,6.1]$ with little variation between different choices for
this range.}
\end{table}
To estimate the systematic error from the small volume,
we have generated a quenched set of $8^4$ and $12^4$ lattices at $\beta=7.77$, $u_0=0.887$
with a similar lattice spacing ($a=0.163(1)$). Analyzing these lattices,
 we find that the $r_0$ extracted from the $8^4$ lattices with
fit range $t\in[3,4]$ is the same within error-bars as the one extracted from the $12^4$ and fit ranges
starting at $3$ or $4$ ranging to $5$ or $6$. 
In the following, we will therefore work with the lattice spacing extracted form the $t\in[3,4]$ fit range.

The dimensionless quantity $r_0\sqrt{\sigma}$ has been used in the past to quantify the
impact of dynamical quarks on the shape of the potential~\cite{Bernard:2001av}.
We find $r_0\sqrt{\sigma} = 1.10(1)$ almost independent of the sea quark mass and $1.18(1)$ for
our quenched ensemble.
On larger lattices and a finer lattice spacing, Ref.~\cite{Bernard:2001av} found a quenched
value of about $1.16$ and 1.128 for two flavors of dynamical staggered quarks. 
From Ref.~\cite{AliKhan:2001tx} we get a value of about 1.14 on at a similar lattice spacing
with two flavors of dynamical clover Wilson fermions. Given the systematic and statistical errors
these values agree well with our findings.

\section{\label{sec:susc}Topological susceptibility}

The topological susceptibility illustrates the strengths and weaknesses of our simulation.
In contrast to all simulations with non-chiral actions, the measurement of the topological charge
in an overlap simulation is trivial. It can even be done during the simulation by
monitoring zero crossings of the smallest eigenmode of the kernel operator
(if the topological charge has been determined once at the beginning of the
simulation).
However, as we have already remarked, the autocorrelation time of the topological
charge during the simulation is annoyingly long.

\begin{figure}
\begin{center}
\includegraphics[width=0.4\textwidth,clip]{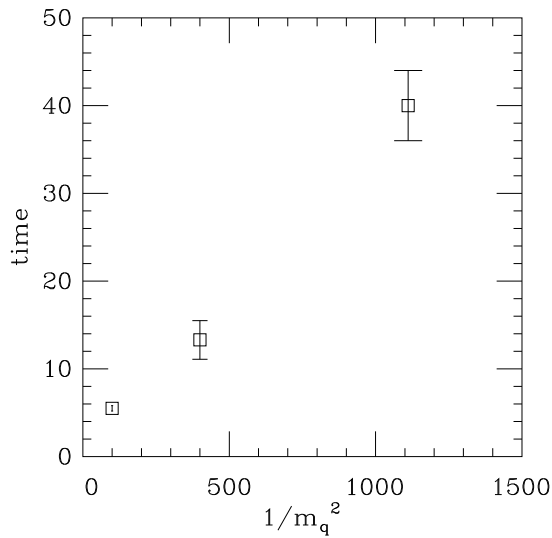}
\end{center}
\caption{Monte Carlo simulation time between topology changes versus quark mass. 
\label{fig:changevmq}
}
\end{figure}

\begin{figure}
\begin{center}
\includegraphics[width=0.5\textwidth,clip]{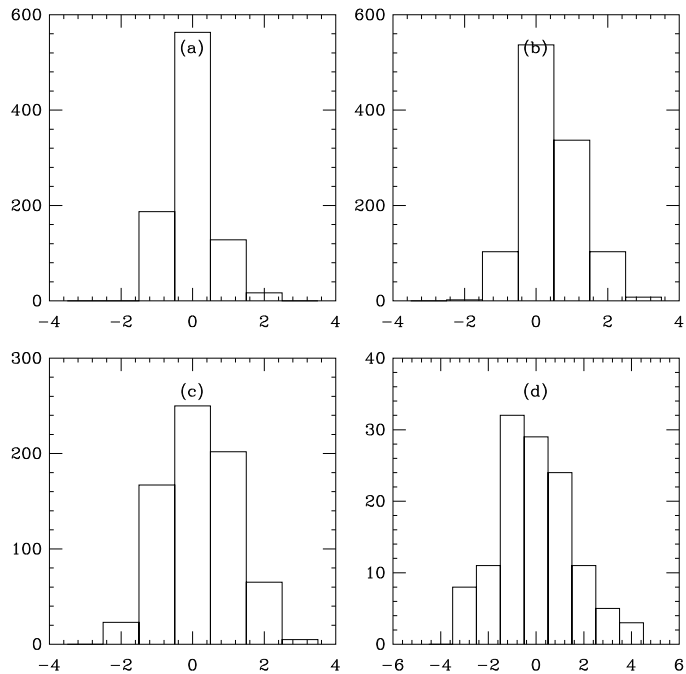}
\end{center}
\caption{Histograms of topological charge from (a) $am_q=0.03$,
(b) $am_q=0.05$, (c) $am_q=0.1$ and (d) quenched simulations.
}
\label{fig:histtopo}
\end{figure}

We begin by showing (in Fig.~\ref{fig:topohist})
time histories of the topological charge for the different simulations
performed at our three values of dynamical quark mass.
Histograms of the topological charge (including
the thermalization runs) are shown in Fig.~\ref{fig:histtopo}.
The topology is recorded at the end of each trajectory (rather than at the end of each time-step).
No sophisticated analysis is needed to see that the autocorrelation time grows
as the quark mass falls, and Fig.~\ref{fig:changevmq} shows that the mean time between
topological changes varies inversely with the square of the quark mass.

With such long times between tunnelings, we are concerned about thermalization effects
in our data. At $am_q=0.1$ $\langle \nu\rangle$ is zero within statistical uncertainty
(it is -0.05(13)
throughout the run) and $\langle \nu^2\rangle$ seems to have stabilized after 50 trajectories
are discarded, so we cut the data there.  At $am_q=0.03$ $\langle \nu\rangle$ is -0.15(11)
when 100 trajectories are discarded from each run and 0.04(12) when 150 are discarded;
$\langle \nu^2\rangle$ shows little variation with cuts of more than 50 initial
trajectories per stream,
 and so we dropped the first 100 trajectories.
At $am_q=0.05$ the situation is similar; we see little variation cutting more than 90-100
configurations and again dropped 100 before averaging.

We find topological susceptibilities of
$\chi a^4 = 2.17(29)\times 10^{-4}$ at $am_q=0.1$,
$\chi a^4 = 1.37(39)\times 10^{-4}$ at $am_q=0.05$,
and
$\chi a^4 = 1.02(24)\times 10^{-4}$ at $am_q=0.03$.

For the quenched data  we are able to space
configurations used in the analysis far enough apart in simulation time as to be
essentially
uncorrelated. We measured a lattice topological susceptibility
 of $\chi a^4=6.13(76)\times 10^{-4}$.
With the quenched $r_0/a=3.08$, this is $\chi\sim(191$ MeV$)^4$,
 which is quite consistent with
typical quenched results, e.g.~\cite{Gattringer:2002mr}.

\begin{figure}
\begin{center}
\includegraphics[width=0.4\textwidth,clip]{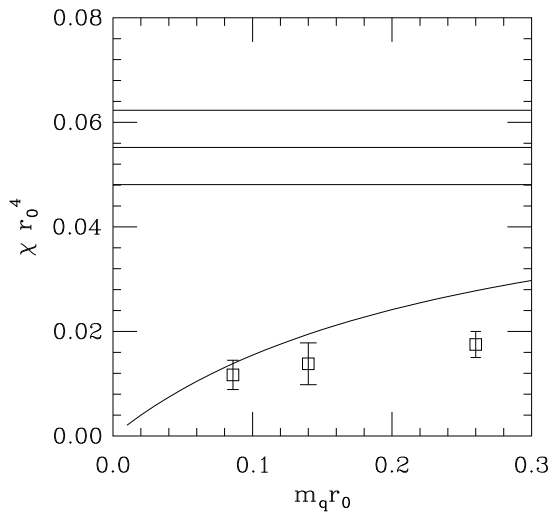}
\end{center}
\caption{Topological susceptibility versus quark mass, in units of
$r_0$. 
The curved line is the D\"urr interpolating formula, Eq. \protect\ref{eq:durr}.
The three horizontal lines give the quenched value and its error.
}
\label{fig:topomr0}
\end{figure}

We attempt to translate this data into dimensionless units in order to facilitate
comparisons with other measurements. We take our measurements of $r_0/a$ from the
previous section to compute $\chi r_0^4$ and do the same with our quark masses,
using the Z-factor as described in the Appendix to convert them to $\mu=2$ GeV
$\overline {MS}$ values.
We present our results in Fig.~\ref{fig:topomr0}.
D\"urr\cite{Durr:2001ty} has presented a phenomenological interpolating formula
for the mass dependence of the topological susceptibility,
in terms of the condensate $\Sigma$ and quenched topological susceptibility $\chi_q$,
\bee
\frac{1}{\chi}=  \frac{N_f}{m_q\Sigma}+ \frac{1}{\chi_q}.
\label{eq:durr}
\ee
Taking $\Sigma$ from our RMT analysis in the next section ( ${r_0}^3 \Sigma=0.43$)
produces the curve shown in the figure.

Lattice results presented elsewhere typically use the pseudo-scalar mass as the
ordinate. As we will describe in the next section, we do not have reliable
pseudo-scalar masses because our lattices are too small. However, we can use the
D\"urr
formula as a benchmark to compare with other simulations.
Our results are in rough agreement in magnitude with those
of another dynamical overlap simulation,
 at lattice spacing
$a\sim 0.25$ fm, by Fodor, et. al.\cite{Fodor:2004wx}. The two simulations both
lie below the D\"urr curve.

Data 
from simulations with two flavors of
ordinary staggered simulations and was analyzed and published
by (among others)
A. Hasenfratz\cite{Hasenfratz:2001wd}. At finite lattice spacing all this data
lies far above the D\"urr curve, and is not too different from the quenched
result.
Other dynamical fermion data with non-chiral fermions\cite{Allton:2004qq}
 is also high with respect to the D\"urr
formula and to overlap data.
It is easy to imagine that simulations with non-chiral actions
would overestimate the topological susceptibility since their massless Dirac operators
do not have exact zero modes.

We do not think that the small volumes
of our simulation have suppressed the susceptibility since our quenched simulations are
not anomalously low.
Little is known about the scaling properties of overlap actions, and our results
and those
of Ref. \cite{Fodor:2004wx} have large lattice spacings.
We are of course not satisfied with the quality of our data from the point of view of
 autocorrelations, lattice spacing, extraction of hadron masses, and simulation volume.

\section{\label{sec:twopt}Meson two-point functions}

Let us now turn to meson two-point functions. The purpose of doing so is two-fold.
First, we want to get an idea of the pion masses at which we are simulating. This
will not work very well since the three volume is small and the time extent is 
far too small for the excited states to decay, but it can provide us with an
upper limit of the pseudo-scalar mass. The second purpose is to
compare the two-point functions to quenched results on similarly sized lattices
and look for effects of the dynamical fermions in the scalar correlator.
We compute zero-momentum correlators 
\bee
C_{ij}(t) = \frac{1}{V} \sum_{\bf x} \langle \bar \psi (0,0) \Gamma_i \psi(0,0) \
\bar \psi ({\bf x},t) \Gamma_j  \psi({\bf x},t) \rangle
\ee
where the fermion fields $\psi$ are contracted with the appropriate flavor structure.
We compute the quark propagators with Gaussian sources of radius $2a$ on gauge configurations
in Coulomb gauge. We use point sinks and apply low-mode averaging using the 
four lowest eigenmodes of the Dirac operator \cite{DeGrand:2004qw,DeGrand:2004wh}.

\begin{figure}
\includegraphics[width=0.3\textwidth,angle=-90,clip]{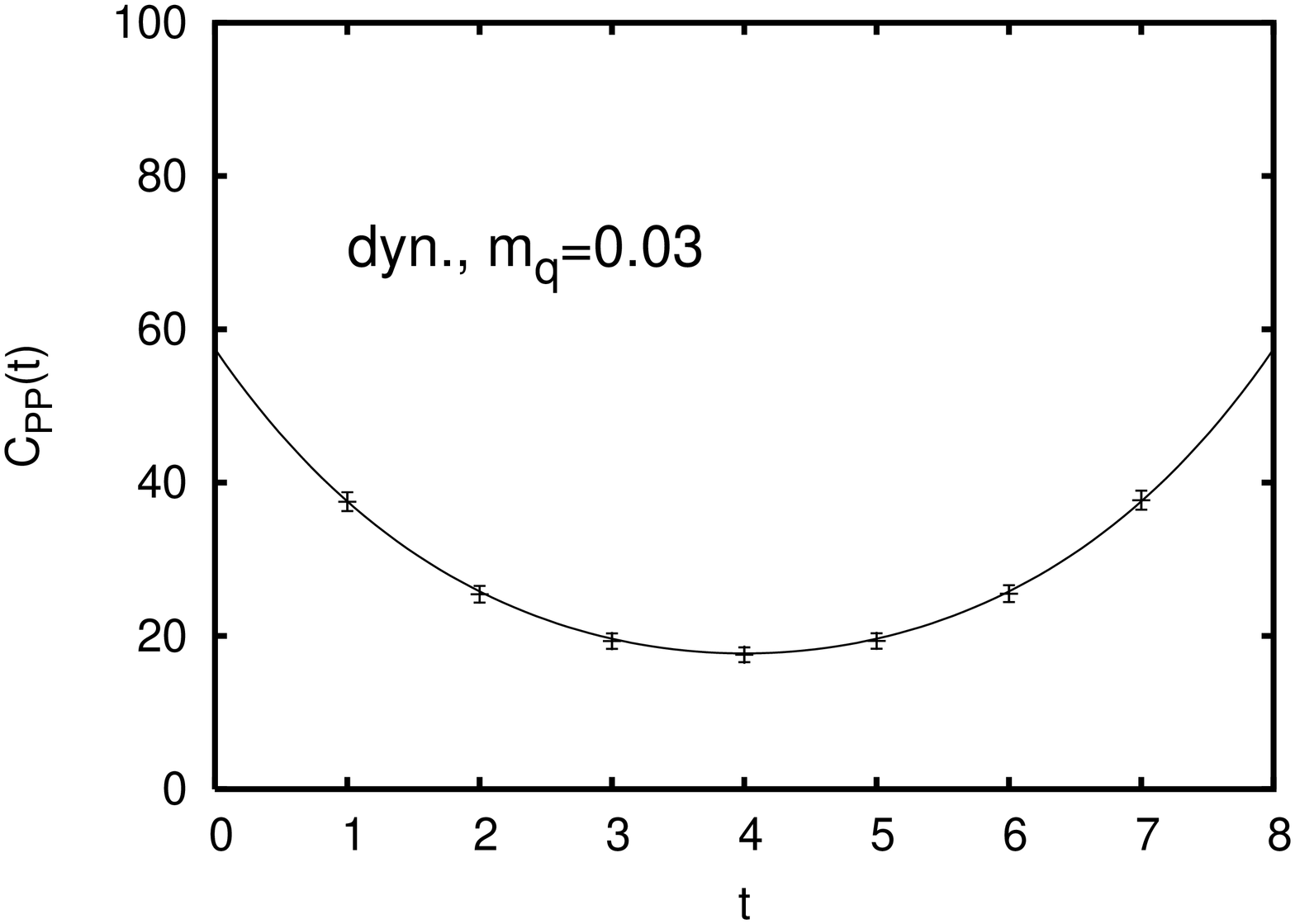}
\includegraphics[width=0.3\textwidth,angle=-90,clip]{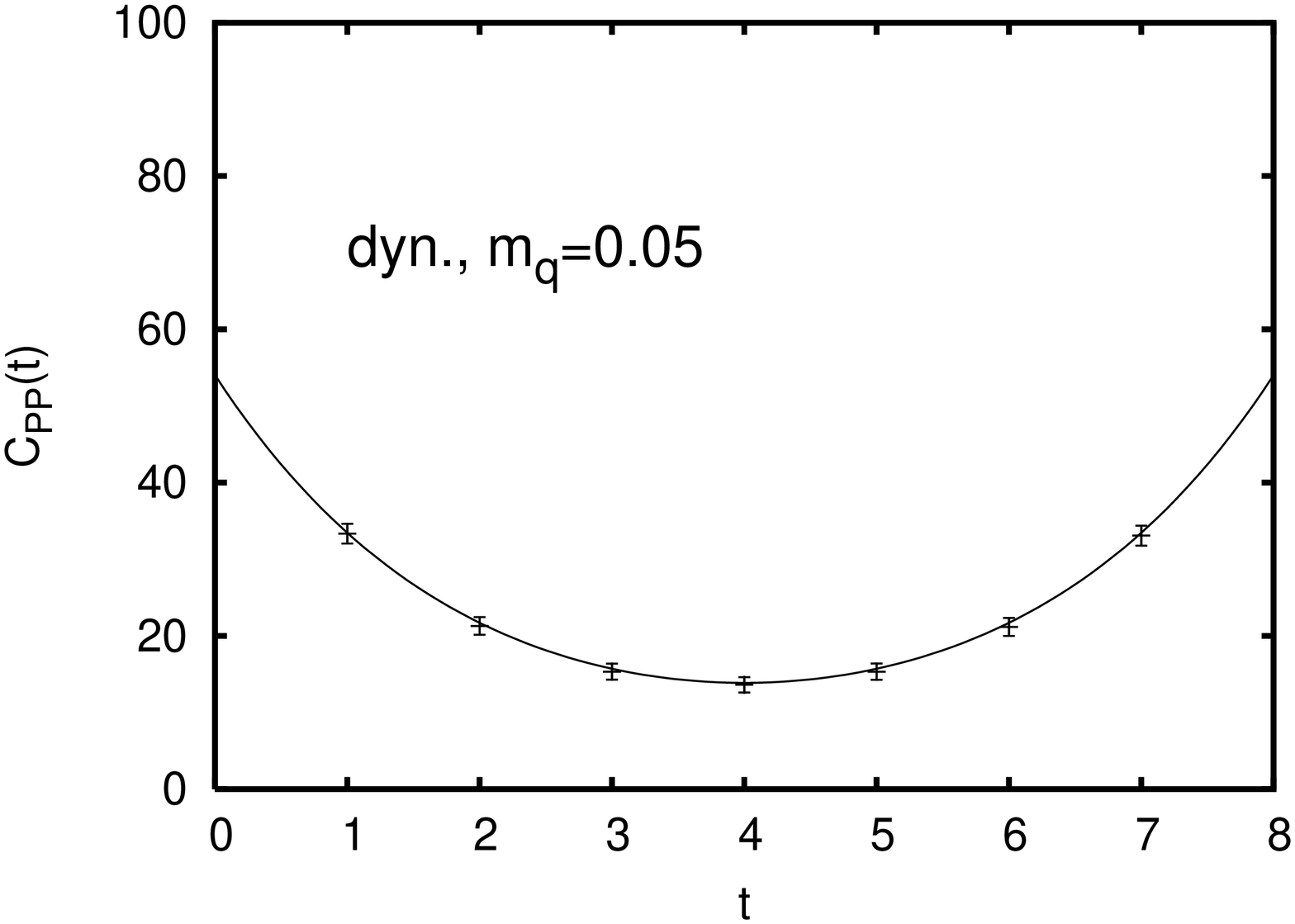}
\caption{\label{fig:pstwopt}The pseudo-scalar zero momentum  two-point function for $m_q=0.03$
and $m_q=0.05$. The lines represent a fit to a single cosh in the range $t\in[2,6]$.}
\end{figure}

\begin{figure}
\includegraphics[width=0.3\textwidth,angle=-90,clip]{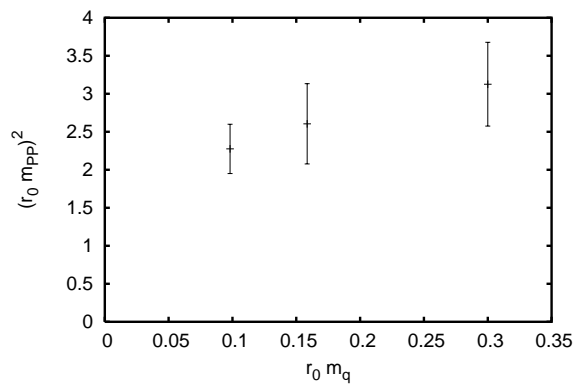}
\caption{\label{fig:psmass}The squared mass of the pseudo-scalar meson as a function of the 
bare quark mass. It does not follow the Gell-Mann--Oakes--Renner relation.}
\end{figure}

In Fig.~\ref{fig:pstwopt} we show the pseudo-scalar two-point function for our two smaller
dynamical quark masses. Unfortunately, the corresponding masses, shown in Fig.~\ref{fig:psmass}, 
do not differ much. This can
be attributed to  the fact that $T=8$ is just too small. We thus see a superposition
of excited states with the ground state which does not depend on the quark mass.

\begin{figure}
\includegraphics[width=0.3\textwidth,angle=-90,clip]{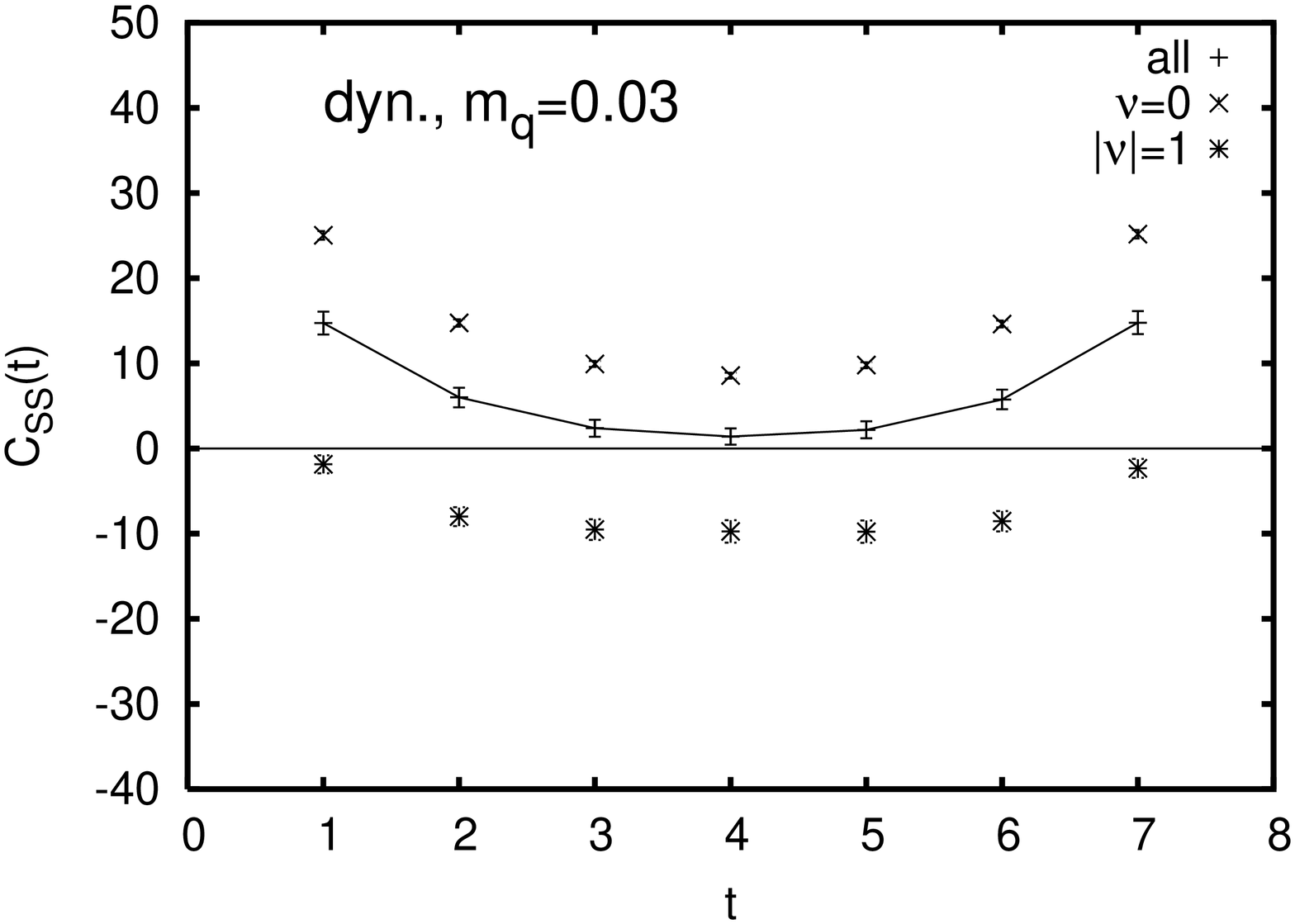}
\includegraphics[width=0.3\textwidth,angle=-90,clip]{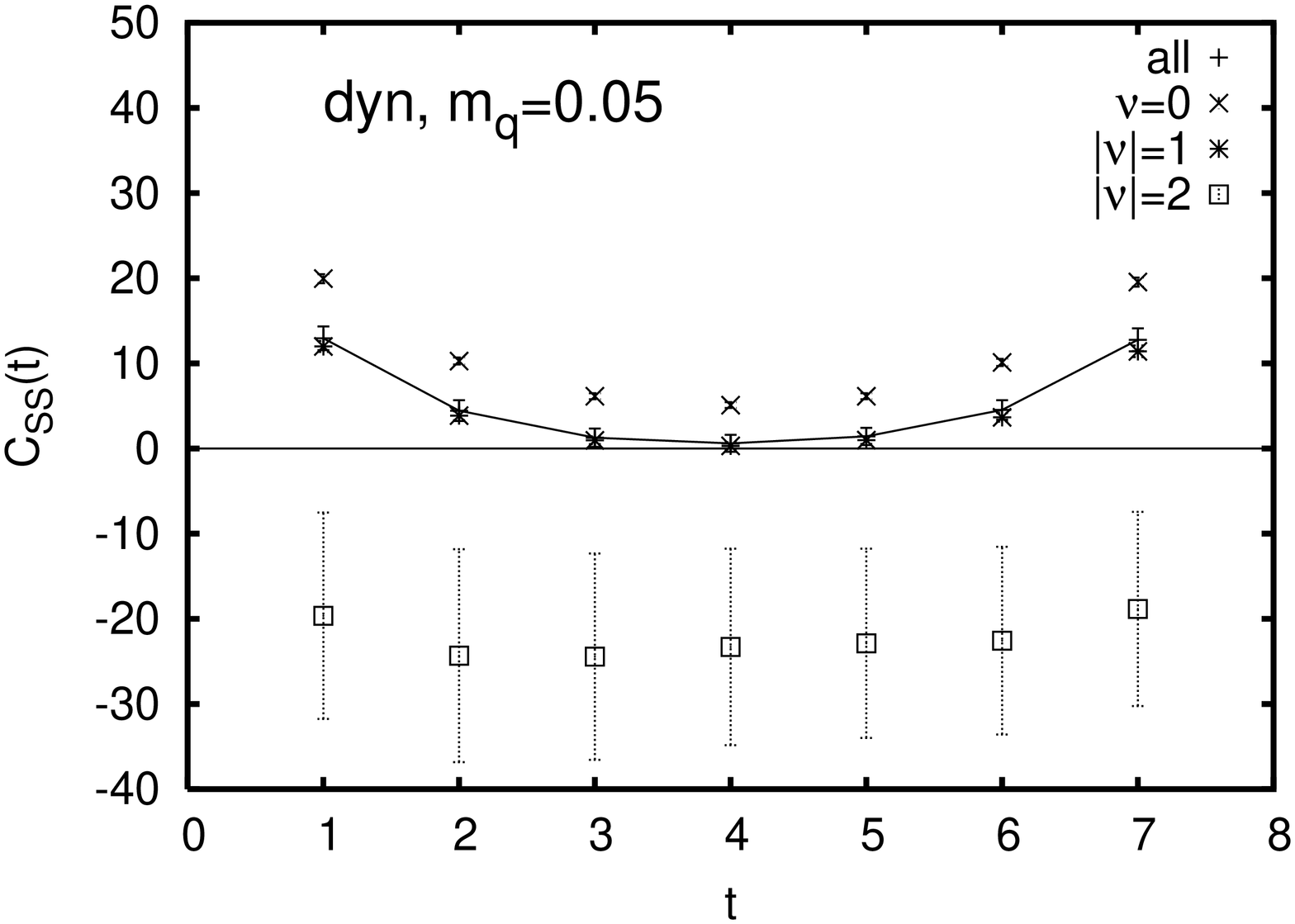}
\includegraphics[width=0.3\textwidth,angle=-90,clip]{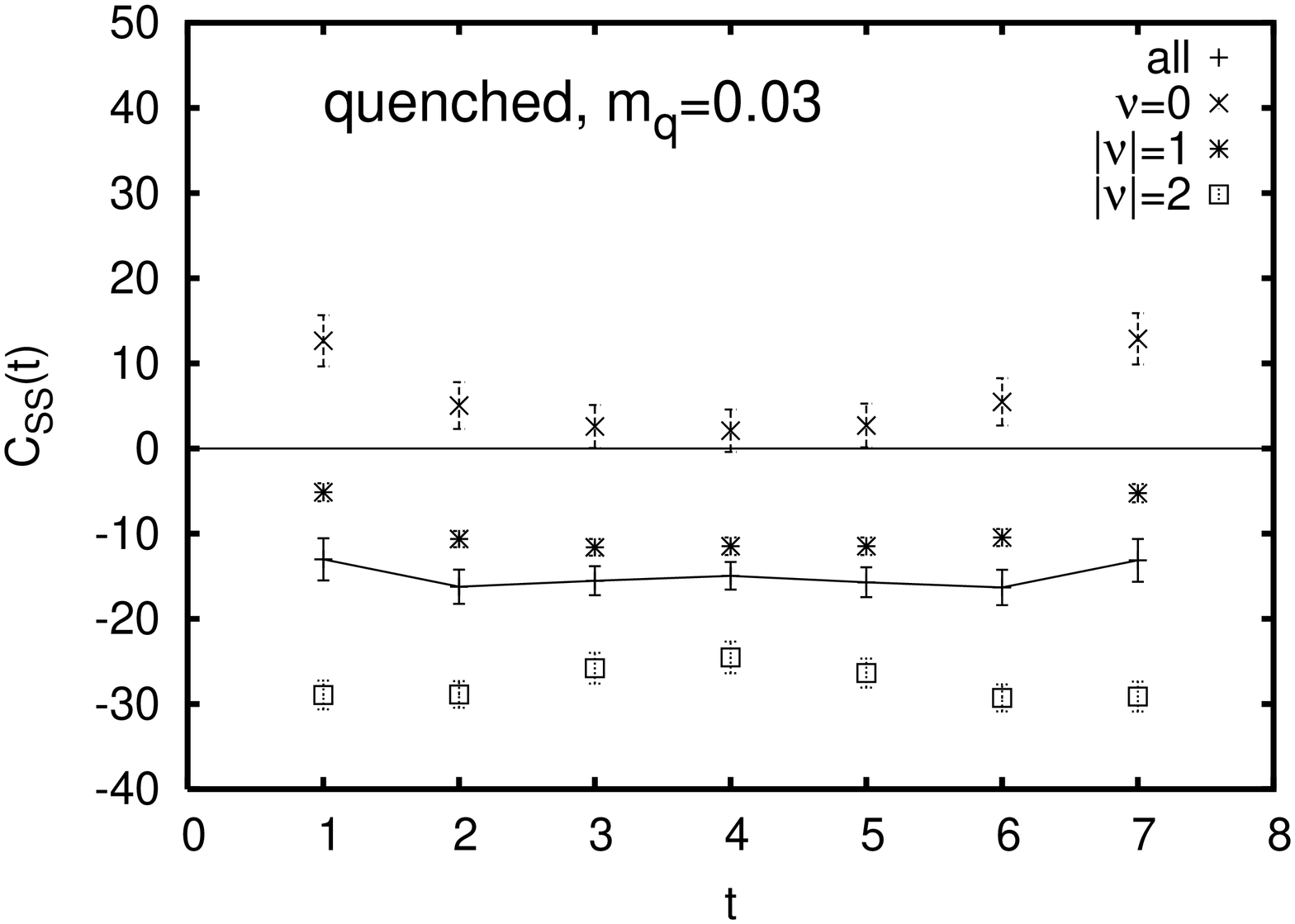}
\caption{\label{fig:stwopt}The scalar zero momentum  two-point function for the 
dynamical $m_q=0.03$ and $m_q=0.05$ ensembles and for the quenched ensemble with
a valence quark mass of $m_q=0.03$. Contrary to the quenched theory, the scalar two-point function stays 
positive with dynamical quarks. We show the full two-point function (connected by a line) and the contributions
from the each topological  sector  separately.  The topologically non-trivial configurations
turn the correlator negative in the quenched theory, where they are more abundant. }
\end{figure}

More impressive is the comparison of the scalar correlator from the dynamical and the quenched
ensemble. It is known that this correlator turns negative in the quenched theory, which
is  a sign that quenched QCD is not  unitary. In Fig.~\ref{fig:stwopt} (bottom) we show
that this is the case even on  a small lattice. The full two-point function is shown (connected
by a line) as well as  the contributions of the various topological sectors. Whereas the function
is positive in $\nu=0$, the zero-modes in the sectors of non-trivial topology turn it
negative. In the full theory, however, the fermion determinant suppresses the sectors
with $|\nu|>0$ sufficiently for the whole two-point function to remain positive 
(Fig.~\ref{fig:stwopt} upper row).

\section{The condensate from Random Matrix Theory}

It was proposed more than a decade ago that the distribution of the low-lying eigenvalues
of the QCD Dirac operator in a finite volume can be predicted by random matrix 
theory (RMT)~\cite{Shuryak:1992pi,Verbaarschot:1993pm,Verbaarschot:1994qf}.
Since then this hypothesis has received impressive support from lattice calculations, 
mainly quenched simulations 
\cite{Berbenni-Bitsch:1997tx,Damgaard:1998ie,Gockeler:1998jj,Edwards:1999ra,Giusti:2003gf},  
but also some dynamical ones using staggered quarks 
\cite{Berbenni-Bitsch:1998sy,Damgaard:2000qt}.

Typically, the predictions are made in the so-called epsilon regime,
 for which $1/\Lambda \ll L \ll 1/m_\pi$ with $\Lambda$ a typical hadronic scale.
However, it has been found that they describe the data in a wider range.
In a recent large scale study, e.g.,  using the overlap operator on quenched 
configurations~\cite{Giusti:2003gf},  it could be shown that from lattices
with a length larger than $1.5~{\rm fm}$, the RMT predictions match the result of the
simulation. Our dynamical lattices have a spatial extent of about $1.3~{\rm fm}$.
As we will see, random matrix theory describes our low-lying Dirac spectra quite well.

Our analysis is based on 
the distribution of the $k$-th eigenmode from RMT as presented
in Ref.~\cite{Damgaard:2000ah} and successfully compared to simulation results
in Ref.~\cite{Damgaard:2000qt}. The prediction is for the distribution of the
dimensionless quantity $\zeta=\lambda_k \Sigma V$ in each topological sector with
$\lambda_k$ is the $k$-th  eigenvalue of the Dirac operator, 
$\Sigma$ the chiral condensate and $V$ the volume
of the box. These distribution are universal and do not depend on additional parameters
other than the number of flavors, the topological charge and the dimensionless 
quantity $m_q \Sigma V$. (Note that topology effects the distributions, in contrast
to the behavior of staggered fermions seen by Ref.~\cite{Damgaard:2000qt}.)
By comparing the distribution of the eigenmodes with the RMT
prediction one can thus measure the chiral condensate $\Sigma$.
The main advantage of this method  is that it gives the
zero quark mass, infinite volume condensate directly. The validity of the approach
can be verified comparing the shape of the distribution for the various modes and topological
sectors. The main uncertainty comes from a too small volume which causes deviations in the
shape, particularly for the higher modes.
As we will see below, the 
direct measurement of $\Sigma$ from $(\bar \psi \psi)(m)$, which requires extrapolation
into the chiral limit as well as correction for the finite volume, is unreliable for 
our data.

\begin{figure}
\includegraphics[width=0.5\textwidth,clip,angle=-90]{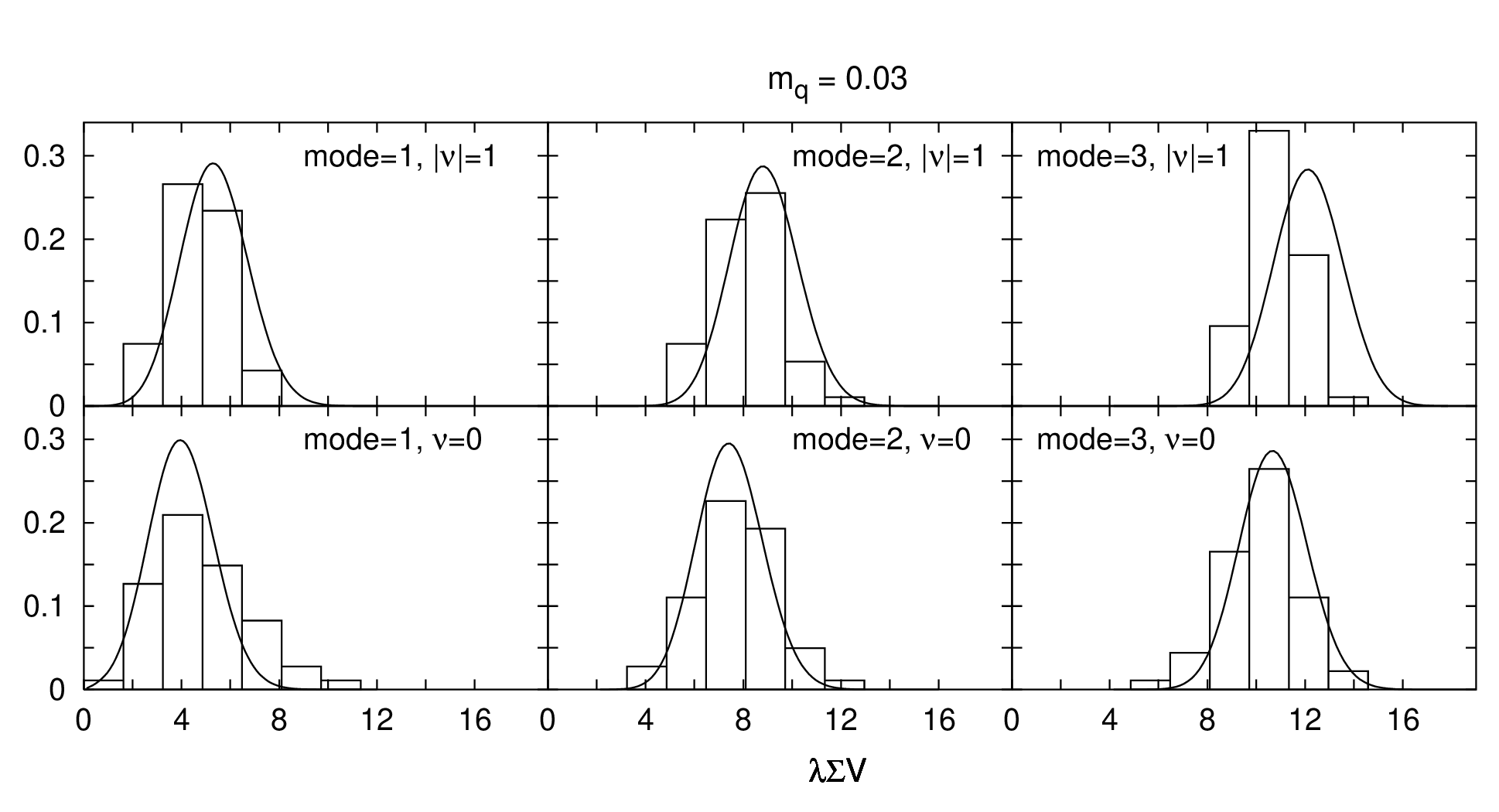}
\includegraphics[width=0.5\textwidth,clip,angle=-90]{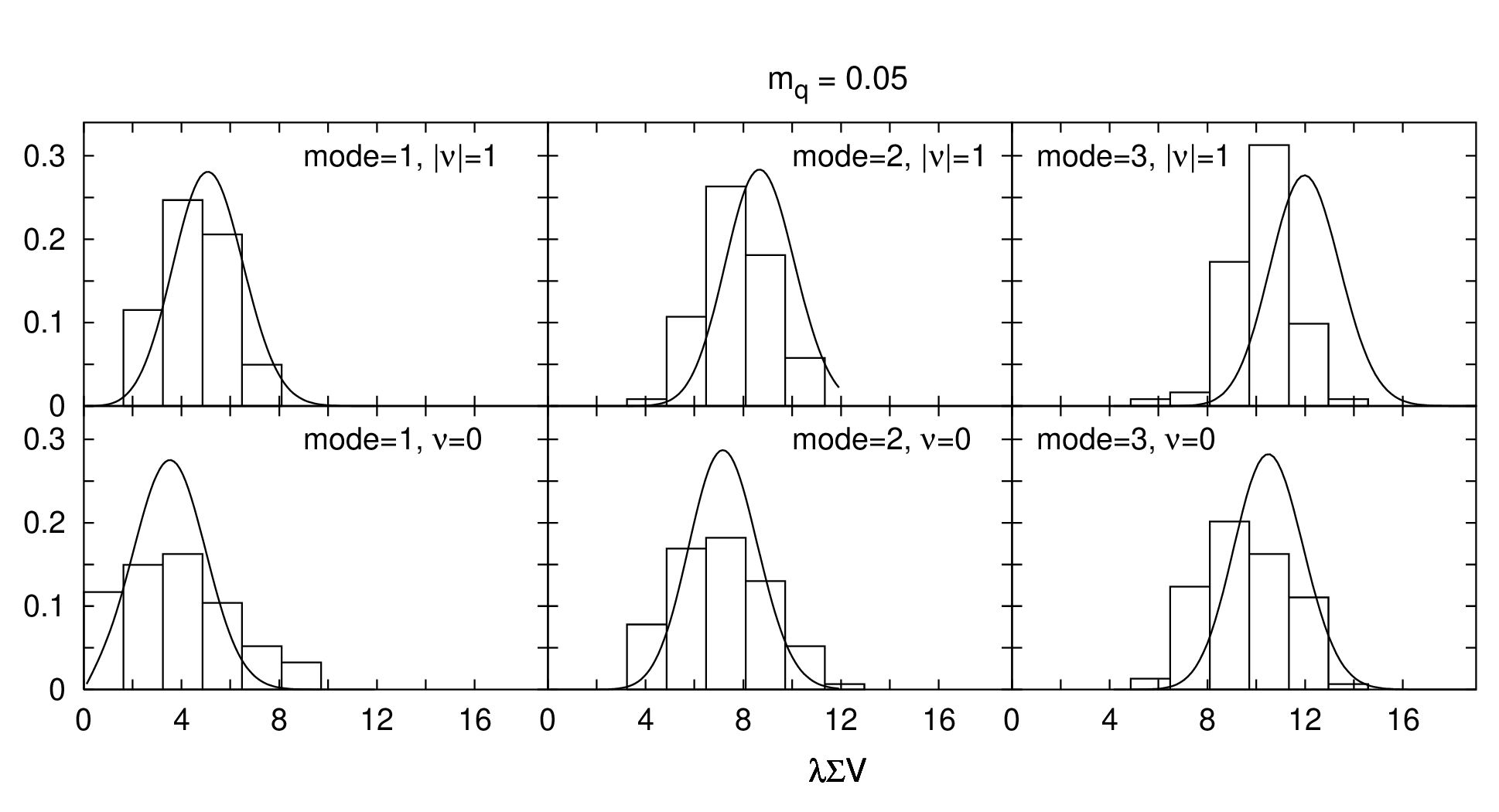}
\caption{\label{fig:rmt}The distribution of the lowest two eigenmodes of the
Dirac operator for our ensemble for the sector of trivial topology and
$\nu=\pm1$. The lines are the result of  fits of the random matrix
theory prediction to the data for the two lowest modes. They correspond to $\Sigma V/a= 54(2)$ and
$\Sigma V/a= 53(2)$ for the $am_q=0.03$  and $am_q=0.05$  ensemble, respectively.
The lines for the third mode are predictions.}
\end{figure}

In Fig.~\ref{fig:rmt} we show the distribution of the two lowest eigenmodes of the overlap
operator (scaled by $\Sigma V$) 
measured on the $\nu=0$ and $\nu=\pm1$ parts of the  $am_q=0.03$ and $am_q=0.05$ ensembles.
To correct for the fact that our eigenvalues lie on a circle of radius $R_0$ 
instead of a straight line, we use the stereographic projection 
$\tilde \lambda=|\lambda|\sqrt{1-|\lambda|^2/2R_0}$.
We fit the RMT prediction from Ref.~\cite{Damgaard:2000ah} to these distributions.
The chiral condensate $\Sigma$ is the only free parameter in this fit; we get 
 $\Sigma V/a= 54(2)$ and $\Sigma V/a= 53(2)$, respectively. This corresponds to
$r_0^3 \Sigma = 0.46(2)$ and 0.41(2).
We studied the dependence of this result on the number of trajectories used in the thermalization
and the separation of consecutive configurations. We found no systematic variation beyond
omitting the first 100 trajectories and separating them by 5.
The prediction agrees overall well with the measured distribution given the low statistics.
However, the distribution of the lowest mode in the $\nu=0$ sector seems to have a tail at larger
$\lambda \Sigma V$ that does not match the prediction. This could be an effect of the small volume.
We also show the prediction for the distribution of the third mode from our fitted values
of $\Sigma V$ in the third column of Fig.~\ref{fig:rmt}. The RMT curve and the data, again, 
agree quite well. However for the $|\nu|=1$ sector, the curve seems to be on the right of the data.
This is probably a sign of the break-down of RMT for eigenvalues larger than the Thouless 
energy~\cite{Osborn:1998nm,Gockeler:1998jj}.

The RMT predictions are made with the assumption that the volume is infinite.
We (obviously) are not in that situation. In finite volume, in the epsilon-regime
of chiral perturbation theory, finite volume modifies the formula for the condensate
by multiplication by a shape factor, $\Sigma \rightarrow \rho\Sigma$,
where
\bee
\rho = 1 + \frac{c(l_i/l)}{f_\pi^2L^2}
\ee
and $c(l_i/l)$ depends on the geometry\cite{Gasser:1986vb}. We do not know $\rho$ since we have
not measured $f_\pi$, but combining our lattice spacing and lattice size with $f_\pi=93$
MeV gives $\rho \sim 1.4$. This is too large a correction to be trustworthy; again, we need
a lattice with a larger physical size.
Our two $r_0^3 \Sigma$ values must be multiplied by
$S_S$ and $\rho$ to give a continuum value: with $r_0=0.5$ fm,
$\rho\Sigma(\overline{MS}) = $ 0.032(1) or 0.029(1) GeV${}^3$
or $(\rho\Sigma(\overline{MS}))^{1/3} =$ 316 or 304 MeV. (Errors are from $r_0/a$
and the fit uncertainty in $\Sigma$.)

\begin{figure}
\includegraphics[width=0.4\textwidth,clip]{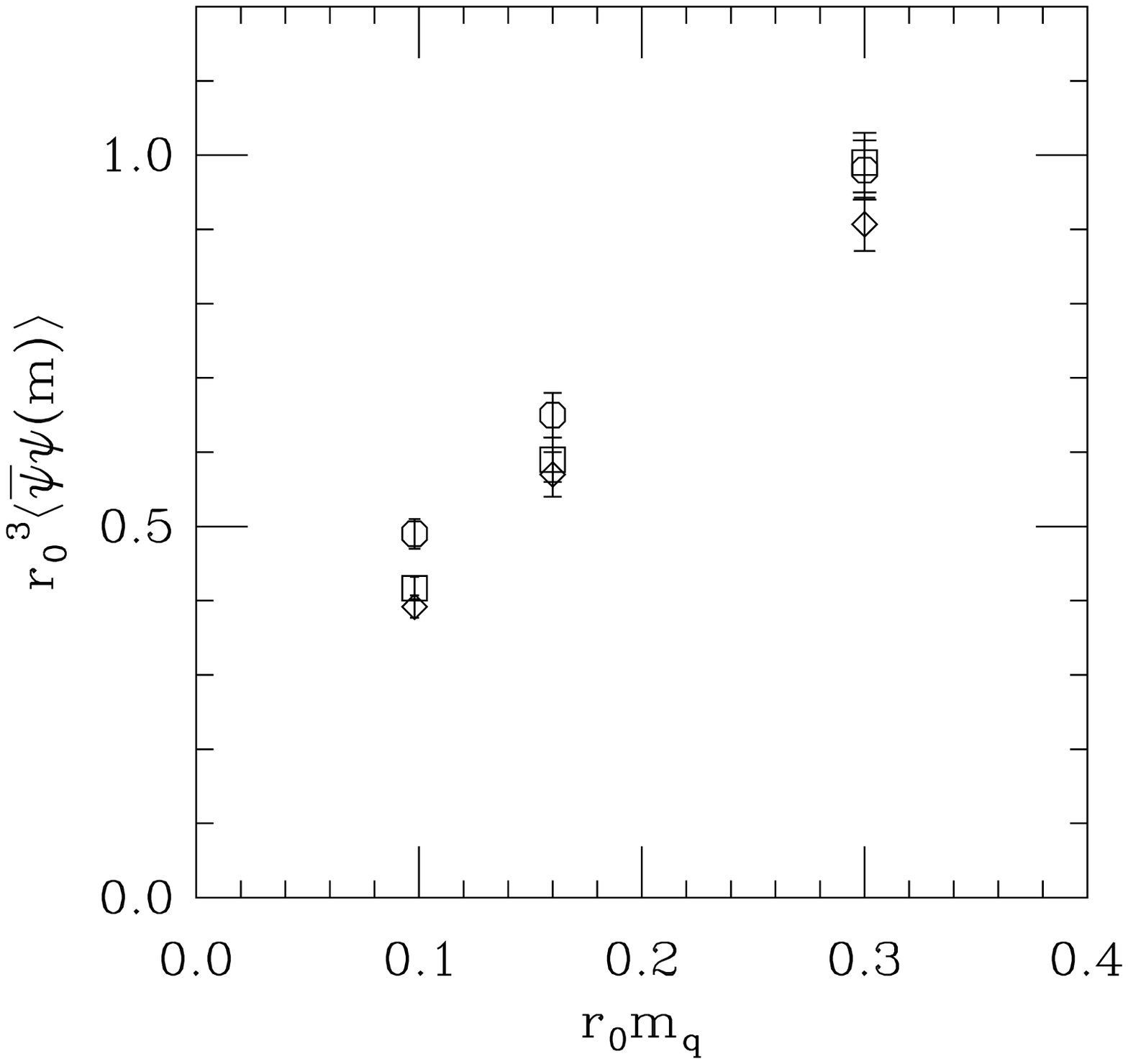}
\caption{\label{fig:sigmad} The mass-dependent condensate vs quark mass,
 scaled by appropriate
powers of $r_0$, from direct measurement. 
 Octagons show the full result (summed over all topological sectors). Squares and diamonds
show $\bar \psi \psi(m)\rangle$ in sectors of topological charge $|\nu|=0$ and 1
 respectively, the latter with the zero-mode
contribution removed.}
\end{figure}

We also attempted to measure the  mass dependent quark condensate
 $\langle \bar \psi \psi(m)\rangle$ using 12 random sources per lattice
on subsets consisting of about 50 lattices per mass value, spaced by
5 trajectories. Our data is displayed in Fig.~\ref{fig:sigmad}.
We show the mass-dependent condensate summed over all topological sectors
as well as its value in sectors of topological charge $\nu=0$ and $|\nu|=1$  (with the
zero mode contribution removed for the latter case). As expected, there is no divergence in
$\langle \bar \psi \psi(m)\rangle$ at zero quark mass (as would occur in a
 quenched simulation).

The parameter $\Sigma$ must be extracted from a fit of $\langle \bar \psi \psi(m)\rangle$
to some functional form. For example, were we in a large volume, we would
fit to the usual constant plus chiral logarithm plus $Cm_q$ formula.

We believe that we are, in fact, in a small volume, and that the appropriate fitting function
is that for the so-called $\epsilon-$regime, given  by
Gasser and Leutwyler\cite{Gasser:1986vb}
for the average-topology case, or by Leutwyler and Smilga\cite{Leutwyler:1992yt}, 
for the case of fixed topology. In either case condensate is
given by
\bee
\Sigma(V,\mu) = \frac{\rho\Sigma }{Z} \frac{ \partial Z}{\partial \mu}
\label{eq:conden}
\ee
with $Z$ the appropriate partition function written by the above authors or
given equivalently by random matrix theory,
and $\mu=m_q\Sigma V$.
We append a linear $Cm_q$ term to Eq. \ref{eq:conden} and attempt to fit the parameters $\Sigma$
and $C$ to the data. This was not successful.

\begin{figure}
\begin{center}
\includegraphics[width=0.4\textwidth,clip]{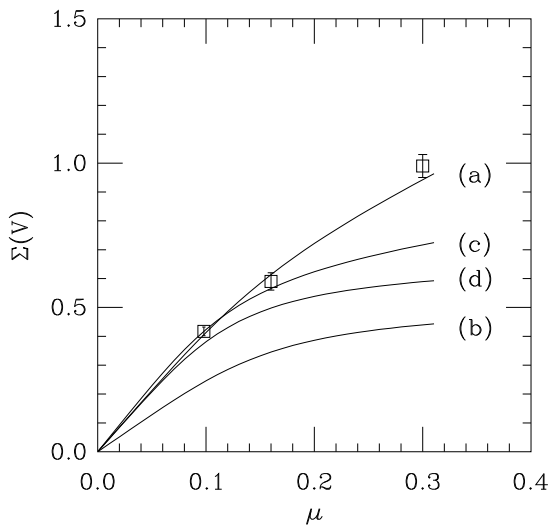}
\end{center}
\caption{Examples of fits to the $\nu=0$ mass-dependent chiral condensate. Curves
(a) and (c) use the three mass points and the two lower ones, respectively.
Curves (b) and (d) show the  part of the fit coming from the
$\Sigma(V,\mu)$ expression of Eq. \ref{eq:conden}.
}
\label{fig:fitnu0}
\end{figure}

First, we do not know which of our mass values lie in the region of validity of our
fitting function.
Fig.~\ref{fig:fitnu0} shows a set of fits to the $\nu=0$ condensate. The fitting function is
a sum of the fixed-topology expression
plus a linear mass-dependent term. Curves (a) and (c) show the result of fits to the
three or lowest two mass data points. Curves (b) and (d) are the part of the fit coming from the
$\Sigma(V,\mu)$ expression of Eq. \ref{eq:conden}.
The three-mass fit gives $r_0^3\Sigma = 0.53(4)$ while the two-mass fit gives 0.67(8).
These numbers are obviously unstable.
We are currently generating data at smaller quark masses which should alleviate this problem
by filling in the curve.

But the numbers we record are quite different from the RMT ones. We think
(see Ref. \cite{Damgaard:2001ep} for a useful discussion) that this is an 
artifact of the small simulation volume.
The problem of principle that we face is that the formulas to which we fit
$\langle \bar \psi \psi(m)\rangle$ assume that the condensate is given by an integral
over a known spectral density $\rho^{(\nu)}(\zeta,\mu)$ for all values of $\zeta$,
\bee
\frac{\Sigma_\nu(\mu)}{\Sigma} = 2 \mu \int_0^\infty d\zeta \frac{\rho^{(\nu)}(\zeta,\mu)}{\zeta^2+\mu^2} \ .
\ee
We are doing simulations in small coarse lattices, and so this assumption is unwarranted.
We expect that cutoff effects will alter the high eigenvalue part of the spectrum.
 Only a larger lattice will cure this problem.

We can check this assumption by fitting $\langle \bar \psi \psi(m)\rangle$ in
the $|\nu|=1$  sector. We get $\Sigma=0.70(3)$ from a fit to the $am=0.03$, 0.05 and 0.1 data sets,
and 0.88(10) from the $am=0.03$ and 0.05 sets. These are bigger discrepancies
from the RMT results than the $\nu=0$ fits.  Fig.~\ref{fig:rmt} shows that the
third eigenmode of the $\nu=0$ sector matches the RMT prediction much better than 
the third $|\nu|=1$ mode does. 

\section{Conclusion}

We have presented results from simulations of two flavors of dynamical overlap fermions
at sea quark masses of about 35, 55 and 100 MeV.
For us, this is a second step in gaining experience with these simulations. We
therefore focused on dynamical fermion effects in the results. We were able to
show that the topological susceptibility is greatly reduced as compared to 
quenched simulations. This measurement also is a great strength of the 
use of overlap fermions for the sea quarks. The topological charge as defined
by the index theorem has a direct impact on the update of the gauge fields during
the simulation. However, we are still worried about the long auto-correlation
time of the topological charge.

We also extracted the quark condensate  from a comparison of the
distribution of the lowest eigenvalues with random matrix theory. This method is
much simpler than a direct fit to $\langle \bar \psi \psi(m)\rangle$ and 
has the advantage of giving the infinite volume, zero mass condensate directly 
without need of extrapolation. The eigenvalue  distributions fit consistently 
the distributions in the various topological sectors for the first three modes. 
However, due to our very small volume, there is a finite size correction to $\Sigma$
which is not well under control. This correction is probably ${\cal O}(40\%)$ for
our simulation. But it is expected to scale which $1/L^2$ and one thus only needs
a moderately larger volume to make it small.

We were also able to demonstrate the effect of the dynamical quarks on 
the scalar meson two-point function. We showed that contrary to the quenched
theory the scalar two-point function is not turned negative by the zero-modes
in the sectors of non-trivial topology. 

All this is possible through an efficient implementation of the overlap operator using
fat links, multiple pseudo-fermion fields to decrease the auto-correlation time
of the topological charge and improvements in the algorithm namely the computation
of the height of the step in the fermionic action.

\appendix
\section{Z-factors from perturbation theory}
A simple application of the techniques described in Ref. \cite{DeGrand:2002va}
gives us one loop predictions
for the vector, axial vector, pseudo-scalar, and scalar currents for our action.
They are shown in  Table \ref{tab:zzz}.
The value of the momentum scale at which the strong coupling constant is evaluated
(from the Lepage-Mackenzie convention\cite{ref:LM}) is also shown.
At one-loop order there is no difference between the
gauge propagator from a tadpole improved action and the tree-level one, so we use the tree-level
propagator in our computation. In this order of perturbation theory, each step of stout smearing
is equivalent to a step of unitarized APE-smearing\cite{ref:APEblock} with
$\rho=\alpha/6$ in the terminology of Ref. \cite{Bernard:1999kc}.

In practice, we define the coupling through the so-called ``$\alpha_V$'' scheme.
We only know the one-loop expression relating the plaquette to the coupling; it is
\bee
\ln \frac{1}{3}{\rm \Tr}U_p = -\frac{8\pi}{3}\alpha_V(q^*) W
\ee
with $W=0.366$ and $q^*a=3.32$ for the tree-level L\"uscher Weisz action. In our calculation of
the condensate, we take the lattice spacing from the Sommer parameter. The coupling
from the plaquette is matched to its $\overline{MS}$ value
and run to the needed value of $q^*$, where we perform the match. Then the $\overline{MS}$
result is run to $\mu=2$ GeV using the usual two-loop formula.
In our simulations $\alpha_V(3.32/a)=0.192$, 0.193, 0.193,
and $Z_s=1.19$, 1.22 and 1.23 for the $am_q=0.03$, 0.05 and 0.10 data sets. Essentially all the
difference from unity comes from the $\overline{MS}$ running from $\mu=1/a$ to 2 GeV.
Using $Z_m=1/Z_S$ gives the $\overline{MS}$ quark masses quoted in the body of the paper.
\begin{table}
\begin{tabular}{|l|c|c|}
\hline
process &\ \ $z_i$\ \ &\ \  $q^* a$\ \ \\
\hline
    $Z_{V,A}$\ \ &\  \  0.54\ \   &\ \ 2.24 \ \ \\
    $Z_{P,S}$\ \ &\ \   2.96\ \   &\ \ 2.85 \ \  \\
\hline
\end{tabular}
\caption{Table of Z-factors and $q^*$'s for  our action, defined so
$Z_i = 1 + z_i g^2(q^*) C_f /(16\pi^2) =1 + z_i \alpha(q^*)/(3\pi)$.
}
\label{tab:zzz}
\end{table}

\section*{Acknowledgments}

This work was supported by the US Department of Energy. It is a pleasure to 
thank A.~Hasenfratz for  conversations and to P.~Damgaard for
extensive correspondence about random matrix theory. Simulations were performed on the
University of Colorado Beowulf cluster. We are
grateful to D. Johnson for its construction and maintenance,
and to D. Holmgren for advice about its design.

\end{document}